\documentclass[pra,twocolumn,superscriptaddress,showpacs,floatfix,longbibliography]{revtex4-2}
\usepackage{mathrsfs,braket}
\usepackage{amssymb, amsbsy, amsmath, latexsym, dsfont, array, layout,
graphicx,mathrsfs,color,ulem,bm}
\usepackage[colorlinks=true,citecolor=blue,urlcolor=blue]{hyperref}

\begin{document}

\title{Non-Hermitian skin effect in a single trapped ion}
\author{Ziguang Lin}
\affiliation{CAS Key Laboratory of Microscale Magnetic Resonance and School of Physical Sciences, University of Science and Technology of China, Hefei 230026, China}
\affiliation{CAS Center for Excellence in Quantum Information and Quantum Physics, University of Science and Technology of China, Hefei 230026, China}
\author{Yiheng Lin}
\email{yiheng@ustc.edu.cn}
\affiliation{CAS Key Laboratory of Microscale Magnetic Resonance and School of Physical Sciences, University of Science and Technology of China, Hefei 230026, China}
\affiliation{CAS Center for Excellence in Quantum Information and Quantum Physics, University of Science and Technology of China, Hefei 230026, China}
\affiliation{Hefei National Laboratory, University of Science and Technology of China, Hefei 230088, China}
\author{Wei Yi}
\email{wyiz@ustc.edu.cn}
\affiliation{CAS Center for Excellence in Quantum Information and Quantum Physics, University of Science and Technology of China, Hefei 230026, China}
\affiliation{Hefei National Laboratory, University of Science and Technology of China, Hefei 230088, China}
\affiliation{CAS Key Laboratory of Quantum Information, University of Science and Technology of China, Hefei 230026, China}


\begin{abstract}
Non-Hermitian skin effect (NHSE) describes the exponential localization of all eigenstates toward boundaries in non-Hermitian systems, and has attracted intense research interest of late. Here we theoretically propose a scheme in which the NHSE significantly impacts the external motion of a single trapped ion through complex spin-motion dynamics.
On the one hand, we show the competition between the NHSE and the coherent Bloch dynamics.
On the other hand, since the NHSE manifests as a non-reciprocal flow in occupied phonon modes, we demonstrate that such dynamics can have potential applications in cooling and sensing.
Our proposal can be readily implemented using existing experimental techniques, and offers a scalable (in terms of the available ions and phonon modes) simulation platform for relevant non-Hermitian physics.
\end{abstract}

\maketitle


\section{\label{introduction}Introduction}

In a general class of non-Hermitian systems, eigenstates become exponentially localized toward boundaries, giving rise to the non-Hermitian skin effect (NHSE)~\cite{PhysRevLett.121.086803, PhysRevLett.123.066404, PhysRevLett.121.136802, PhysRevB.99.201103, PhysRevLett.121.026808, PhysRevX.8.041031, PhysRevB.97.121401, PhysRevLett.125.126402, PhysRevX.8.041031, PhysRevB.97.121401, PhysRevLett.125.126402, PhysRevLett.124.086801, PhysRevResearch.1.023013, PhysRevLett.125.226402, PhysRevB.100.035102, Nature.communications.11.1.(2020), 2111.04196, 2111.04220}.
The NHSE originates from the spectral topology of the system's complex egienspectrum under the periodic boundary condition, which translates to a non-reciprocal bulk flow that leads to the accumulation of eigenstates at boundaries~\cite{PhysRevLett.125.126402,PhysRevLett.124.086801}. In lattice systems with NHSE, the deviation of eigenstates from extended Bloch waves necessitates the application of the non-Bloch band theory, which offers a consistent and efficient description for topological edge states~\cite{PhysRevLett.121.086803, PhysRevLett.123.066404, PhysRevLett.121.136802} and system dynamics~\cite{PhysRevResearch.1.023013, PhysRevResearch.3.023022, PhysRevLett.123.170401, PhysRevB.102.201103, PhysRevLett.124.066602}.
So far, signatures of the NHSE have been observed in topoelectrical circuits~\cite{Nat.Phys.16.747}, metamaterials~\cite{doi:10.1073/pnas.2010580117}, photonics~\cite{Nat.Phys.16.761,lightfunnel}, and cold atoms~\cite{PhysRevLett.129.070401}. A wealth of unconventional phenomena predicted by the non-Bloch band theory, such as the non-Bloch parity-time symmetry~\cite{PhysRevResearch.1.023013,nonblochep}, non-Bloch topological invariants and quench dynamics~\cite{PhysRevResearch.3.023022,nonblochquench}, have also been experimentally confirmed.
The identification of these NHSE-related behaviors in the quantum mechanical setting such as cold atoms is particularly interesting, as it paves the way for exploring, in quantum open systems, exotic many-body phenomena that can be conveniently understood from the perspective of non-Hermitian physics.

In a recent experiment, dynamic signatures of the NHSE have been observed in a dissipative Bose-Einstein condensate moving along a synthetic momentum lattice~\cite{PhysRevLett.129.070401}. Therein, the implemented model is the dissipative Aharonov-Bohm (AB) chain, consisting of a series of rings with on-site loss and threaded by synthetic flux~\cite{PhysRevLett.124.070402}. However, due to experimental limitations, the implemented system is confined to five unit cells.
Since the NHSE persists in the thermodynamic limit, it is desirable to consider alternative designs and systems where the lattice size can be further increased.

In this work, we propose one such realization using a single trapped ion~\cite{doi.org/10.1038/nphys2252, doi.org/10.1038/nature09801}. Taking the ion of $^{9}$Be$^{+}$ as a concrete example, we demonstrate that a dissipative AB chain can be realized in the combined synthetic dimensions of the hyperfine states and phonon modes. Here a state-selective dissipation can be implemented, by coupling the ground hyperfine state to an excited state undergoing spontaneous decay. Under post selection, the resulting dissipative AB chain is a semi-infinite one-dimensional lattice with a natural open boundary at $n=0$, where $n$ labels the Fock space of the phonon mode.
While the natural open boundary would necessitate the non-Bloch band theory to account for the system's topological edge states, we focus on the dynamic consequence of the NHSE. Specifically, a unidirectional flow emerges in the occupied phonon modes, manifesting itself as increasing or decreasing average phonon numbers, depending on the parameters.
By tuning the system parameters such that the AB chain becomes tilted in the synthetic dimension, the competition between the NHSE and the coherent Bloch dynamics can be studied.
Further, in the regime featuring decreasing average phonon number, the NHSE-induced flow gives rise to an effective cooling of the ion. We also find that the change in the average phonon number is sensitive to the synthetic flux, which is potentially useful for quantum sensing.
Finally, we demonstrate that the directional flow is robust again various experimental imperfections, such as heating and spontaneous decay back into the ground states.

\begin{figure*}[t]
\includegraphics[width=13.6cm]{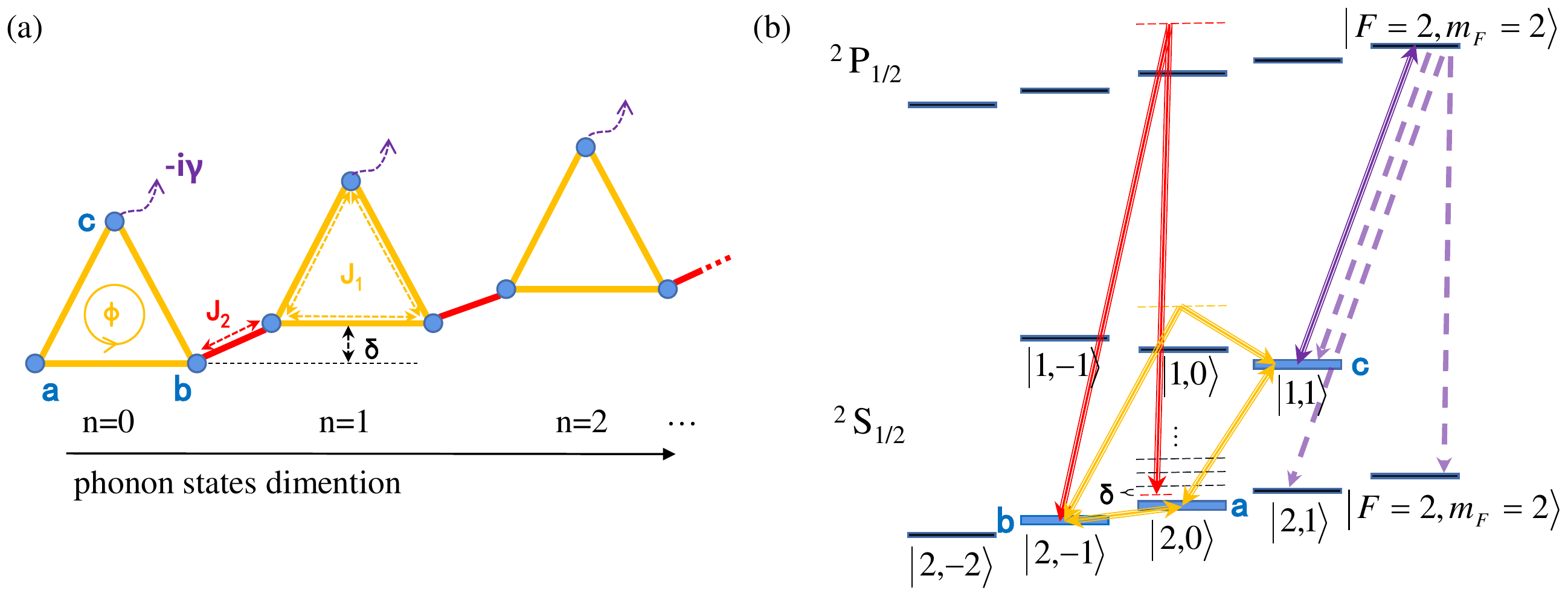}
\caption{ Experimental scheme with a single trapped $^9$Be$^+$ ion. (a) A tilted dissipative Aharonov-Bohm (AB) chain. (b) The level scheme of our proposal. The sublattice sites $(a,b,c)$ in (a) are respectively encoded in the states $(\ket{F=2,m_F=0},\ket{F=2,m_F=-1}, \ket{F=1,m_F=1})$ of the
ground-state $^{2}S_{1/2}$ manifold.
Within each unit cell, sites $(a, b, c)$ are coupled through the carrier ($\Delta n=0$) transitions driven by a radio-frequency field (between $a$ and $b$), a microwave field (between $a$ and $c$) or a combination of both in a two-photon process (between $b$ and $c$).
Their phases contribute to a synthetic magnetic flux $\phi$ through each ring.
Site $a$ of the $(n+1)$th cell and site $b$ of the $n$th cell are coupled through a two-photon Raman transition, resonant with the first sideband ($\Delta n=1$). Thus different unit cells are encoded into the phonon modes. When the Raman transition is detuned by $\delta$, a cell-dependent offset potential is introduced, giving rise to a tilted AB chain.
The laser-induced dissipation is introduced by coupling site $c$ to the excited state $\ket{F=2,m_F=-2}$ of the $^{2}P_{1/2}$ manifold, which spontaneously decays back into the ground-state manifolds. We first neglect the spontaneous decay back into site $c$, but will explicitly gauge its impact in Sec.~V.
The states $\ket{F=2,m_F=1}$ and $\ket{F=2,m_F=2}$ then serve as the reservoir, rendering the dynamics within $(a,b,c)$ non-unitary upon post selection.
}
\label{Fig1}
\end{figure*}

The work is organized as follows. In Sec.~II, we present in detail the proposed scheme, as well as the implemented model. In Sec.~III, we discuss the unidirectional flow in the occupied phonon modes, as a direct consequence of the NHSE. We also depict the competition between the dynamic signatures of the NHSE and the Bloch dynamics. In Sec.~IV, we show the dynamic features of the NHSE can have potential applications in cooling and sensing. In Sec.~V, we demonstrate the robustness of the NHSE under experimental imperfections. We summarize in Sec.~VI.

\section{\label{model}Experimental scheme and model}
As illustrated in Fig.~\ref{Fig1}(a), we aim to simulate dynamics along a dissipative AB chain, consisting of a series of AB rings with synthetic magnetic flux and on-site dissipation~\cite{PhysRevLett.124.070402}. To realize this model, the ability to control coherent couplings and dissipation in a multi-level open system is indispensable.
Implementing a long lattice with a long coherence time is also important to investigate the system dynamics, which in general plays a crucial role in identifying the NHSE and topological properties of the model~\cite{Nat.Phys.16.761,PhysRevLett.129.070401}. A further requirement for the investigation of the NHSE is the ease with which to implement sharp boundaries, which is not the case, for instance, for cold atoms trapped in an optical lattice potential.

To satisfy all these requirements, we propose an experimentally feasible scheme with a single trapped ion. As a concrete example, we consider the $^{9}$Be$^{+}$ ion, whose rich level structure offers a particularly convenient platform. As shown in Fig.~\ref{Fig1}(b), the ground manifold $^{2}S_{1/2}$ of $^{9}$Be$^{+}$ features
eight hyperfine levels due to the coupling between the electronic and nuclear spins ($I=\frac{3}{2}$), with a hyperfine splitting of $\sim 1.2$ GHz. The transitions among these states can be coherently driven by radio-frequency or microwave fields for single photon transitions, and laser fields for Raman transitions. The ground-state manifold is therefore a qudit that can be
well controlled and manipulated. To simulate the dissipative AB chain, we encode the sublattice sites $(a,b,c)$ of the AB rings into the hyperfine states $\ket{F=2,m_F=0}$, $\ket{F=2,m_F=-1}$ and $\ket{F=1,m_F=1}$, respectively. Different unit cells are encoded into different phonon states $\ket{n}$ ($n=0,1,2...$).

While the intracell hoppings are generated by radio-frequency and/or microwave fields, the intercell hopping between $\ket{n+1,a}$ and $\ket{n,b}$ is realized by a stimulated Raman two-photon process, marked red in Fig.~\ref{Fig1}(b). Thus, the spin and motion are coupled along the direction of the Raman lasers.
The Rabi frequencies of these sideband transitions vary with $n$ and are proportional to $\sqrt{n+1}$ if the Lamb-Dicke criterion is satisfied. However, the light mass of $^{9}$Be$^{+}$ leads to a large Lamb-Dicke parameter, such that the Rabi frequencies for $n\leq 14$ are more or less uniform. We therefore first consider a uniform coupling rates, unless othrewise specified (in Sec.~IV).

To realize non-Hermiticity, dissipation is introduced by coherently exciting atoms in the state $^{2}S_{1/2}\ket{F=1,m_F=1}$ to $^{2}P_{1/2}\ket{F=2,m_F=2}$, which has a natural linewidth of $\Gamma=2\pi\times 19.4$~MHz. Upon spontaneous decay, the excited state $^{2}P_{1/2}\ket{F=2,m_F=2}$ has a probability of $1/2$ ending up in either
$\ket{F=2,m_F=1}$ or $\ket{F=2,m_F=2}$ of the ground-state manifold, which are then post-selected. While the ion also has a significant probability ($1/2$) to decay back into the state $\ket{F=1,m_F=1}$ (site $c$) causing decoherence, such a process does not qualitatively change the dynamic signatures of the NHSE, as we will show in Sec.~V. In the following, we first neglect these processes for the convenience of discussion.

Under the condition of post selection, the dynamics of the system is then driven by the non-Hermitian effective Hamiltonian (see Appendix for a detailed derivation)
\begin{equation}
\begin{aligned}
H &=\sum_n n\delta(|n,a \rangle\langle n,a|+|n,b \rangle\langle n,b| +|n,c \rangle\langle n,c|)\\
&-\sum_n i\gamma |n,c \rangle\langle n,c|+ \sum_n[J_1 (e^{i\phi}|n,c \rangle\langle n,a|\\
&+|n,c \rangle\langle n,b|+|n,a \rangle\langle n,b|)+ J_2|n+1,a\rangle\langle n,b|+H.c. ]
\end{aligned}\label{eq:H}
\end{equation}
where $\gamma =J_e^2 /\Gamma$ is the effective dissipative rate, and $J_e$ is the
coupling strength between site $c$ and the excited state $^{2}P_{1/2}\ket{F=2,m_F=2}$. $J_1$ and $J_2$ are respectively the intracell and intercell hopping rates. The tunable phase $\phi$ corresponds to the synthetic magnetic flux through the rings. Note that a unit-cell dependent detuning $\delta$ can be created by detuning the sideband drive from resonance, as shown in Fig.~\ref{Fig1}(b). The most general form of our implemented model is therefore a tilted dissipative AB chain in the synthetic dimensions of the trapped ion.
For $\delta=0$, Hamiltonian (\ref{eq:H}) reduces to the dissipative AB chain, manifesting the NHSE and non-Bloch band topology. Whereas for $\delta\neq 0$, the Hamiltonian resembles a Wannier-Stark ladder~\cite{PhysRev.117.432}, wherein the coherent Bloch dynamics competes with the NHSE.

Since the states of a trapped ion can be detected with very high efficiency and fidelity, dynamics in the synthetic dimension can be conveniently probed. For instance,
the ion's internal states can be read out through fluorescence by coupling to an excited state. The external phonon states can be mapped onto the internal degrees of freedom of the ion via a sideband drive for readout. This is because
sideband transitions between $\ket{n}$ and $\ket{n+1}$ states have different Rabi frequencies with different $n$. By fitting the resultant internal state population after driving the sideband transitions, the population of different phonon states can be reconstructed~\cite{Nature.Communications.12.1126.(2021)}.
In this way, phonon modes up to $n\sim 100$ can be manipulated and probed~\cite{Nature.572.86(2019)}.
This lends a valuable scalability to the dissipative AB chain, whose prior implementation is limited to five unit cells in cold atoms~\cite{PhysRevLett.129.070401}.

\section{\label{dynamic}Dynamic signatures of the NHSE}
The NHSE originates from a directional bulk flow~\cite{PhysRevLett.125.126402}, thus offering a useful dynamic signature for experimental detection. In this section, we demonstrate that this is also the case with our proposed setup.

\subsection{\label{nhse} The NHSE from dynamics}
\begin{figure}[t]
\includegraphics[width=8.6cm]{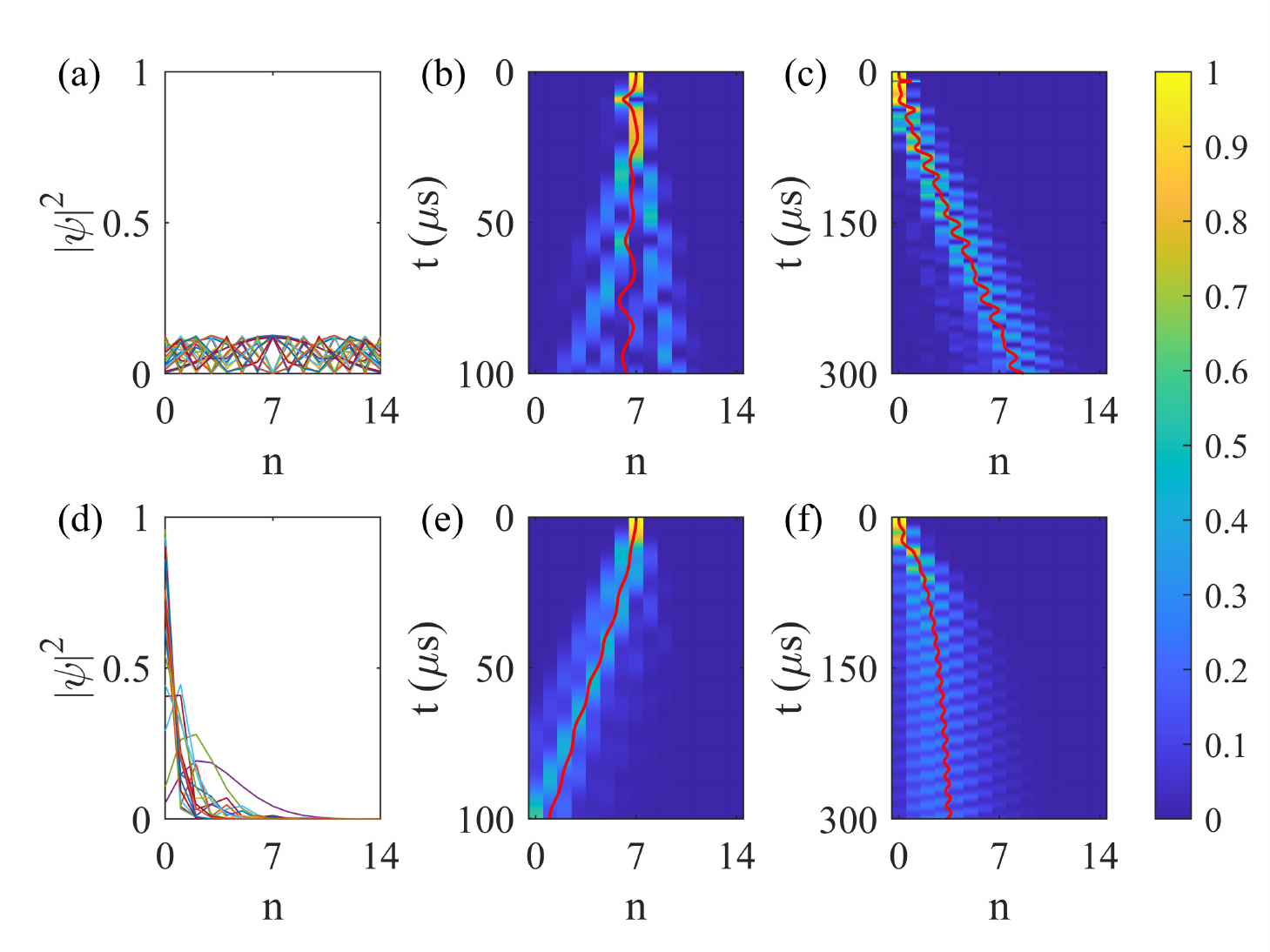}
\caption{ NHSE of the AB chain. (a)(b)(c) are the Hermitian cases with $\gamma = 0$; (d)(e)(f) are the non-Hermitian cases with $\gamma = 2\pi\times50 $~kHz. (a)(d) show the spatial distribution of eigenstates wavefunctions. (b)(e) are the bulk dynamics with an initial state $\ket{n=7,a}$. (c)(f) are dynamics close to the boundary at $n=0$, with the initial state $\ket{n=0,a}$. The color bar indicates the normalized probability in each cell. The red solid line indicates the time evolution of the average phonon number. Other parameters are $J_1 = 2\pi\times20$~kHz, $J_2 = 2\pi\times10$~kHz, $\phi = -\pi/2$ and $\delta = 0$.}
\label{Fig2}
\end{figure}

We first focus on the case with $\delta=0$. The NHSE of the model is then closely related to the sign and magnitude of the synthetic flux $\phi$.
Since in our case, a natural open boundary exists at $n=0$, the NHSE in the synthetic dimension is readily visible from the eigenstates' wavefunctions. As shown in
Fig.~\ref{Fig2}(a)(d), whereas the eigenstates are extended for $\gamma=0$, they become localized toward the open boundary at $n=0$ under finite dissipation. Note that we fix $\phi=-\pi/2$ for the calculations here, since the NHSE requires a finite $\gamma$ and $\phi\neq 0,\pi$.

In Fig.~\ref{Fig2}(b)(e), we show the bulk dynamics for the initial state $\ket{n=7, a}$, either without or with dissipation. A unidirectional flow is clearly visible in Fig.~\ref{Fig2}(e), confirming the emergence of NHSE therein. More directly, the NHSE can be identified through dynamics close to the open boundary. In Fig.~\ref{Fig2}(c)(f), we show dynamics for the initial state $\ket{n=0,a}$. Under the impact of the NHSE, the time-evolved state remains close to the open boundary in
Fig.~\ref{Fig2}(f).

\subsection{\label{bloch}Competition with the Bloch dynamics}
Under a finite $\delta$, Hamiltonian (\ref{eq:H}) corresponds to a tilted AB chain. In the absence of dissipation ($\gamma=0$), eigenstates of the system are localized, since the system essentially constitutes a Wannier-Stark ladder, which is known for its localized eigenstates. While the localization occurs in the bulk, it should compete with the NHSE in the presence of dissipation. Dynamically, whereas the NHSE is signaled by the unidirectional propagation, a Wannier-Stark ladder is noted for the coherent Bloch oscillation~\cite{BO}. We therefore expect the competition between the two distinct types of dynamics to emerge.

\begin{figure}[t]
\includegraphics[width=8.6cm]{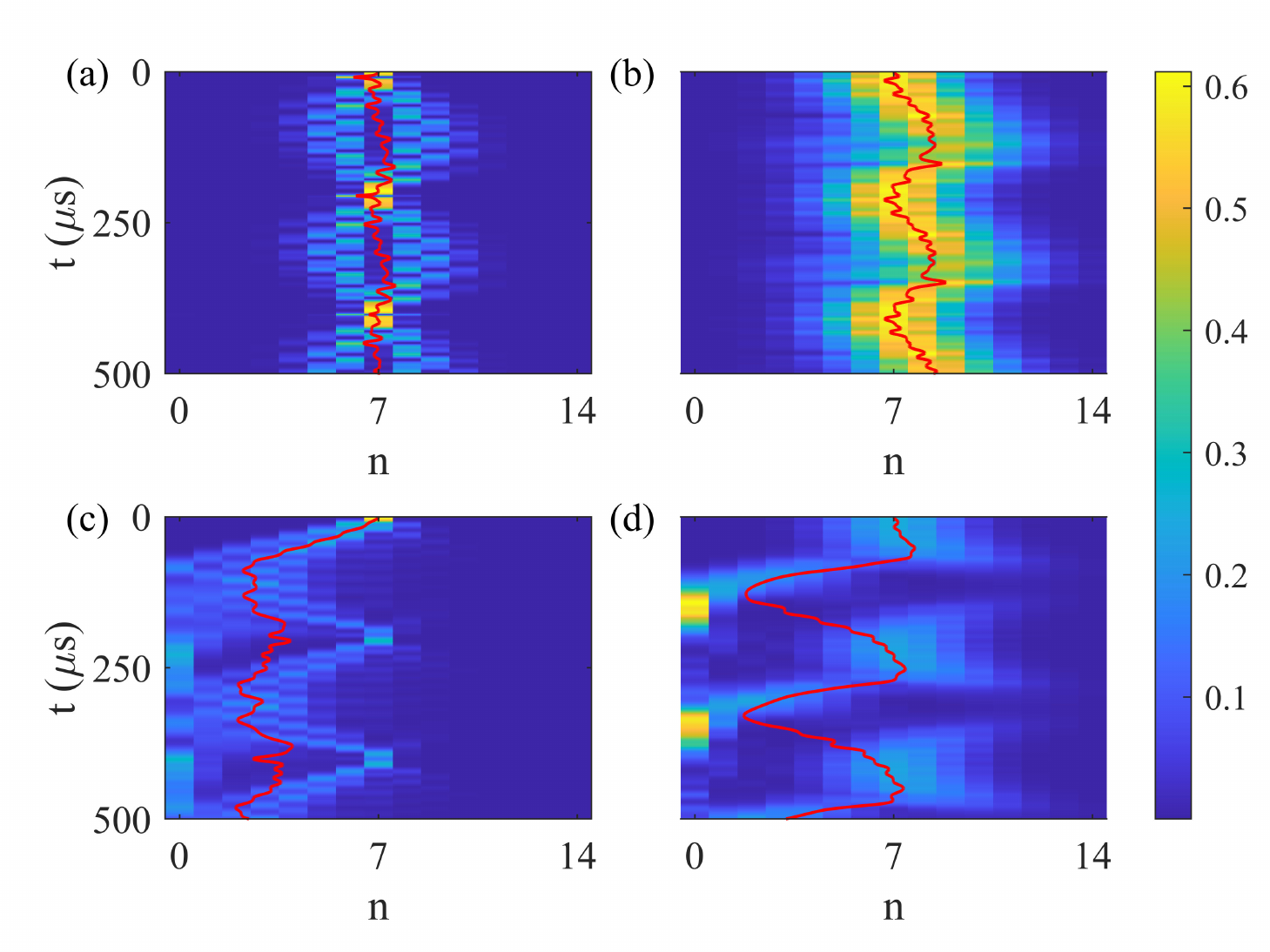}
\caption{ Competition between the NHSE and the Bloch dynamics, where the color bar indicates the phonon number in each unit cell, at a given time.
(a)(b) are the Hermitian cases with $\gamma = 0$; (c)(d) are the non-Hermitian cases with $\gamma = 2\pi\times50$~kHz. The system is initialized in the state $\ket{n=7,a}$ for (a) and (c), and in a Gaussian states $\sum_n e^{-0.1(n-7)^2}\ket{n}$ for (b) and (d). The red solid line indicates the average photon number. Other parameters are $J_1 = 2\pi\times20$~kHz, $J_2 = 2\pi\times10$~kHz, $\phi = -\pi/2$ and $\delta = 2\pi\times5$~kHz.}
\label{Fig3}
\end{figure}

This is confirmed in Fig.~\ref{Fig3}. In the Hermitian case, the Bloch dynamics manifests in two distinct scenarios. For an initial state localized within a single unit cell, the state is close to an eigenstate of the tilted AB chain, such that the average phonon number (red curve) oscillates close to the initial phonon number with a small amplitude [see Fig.~\ref{Fig3}(a)]. By contrast, when the initial state features a wave packet spanning several unit cells, the average phonon number significantly oscillate around the initial value in the dynamics [see Fig.~\ref{Fig3}(b)]. More importantly, in Fig.~\ref{Fig3}(c)(d), we plot the dynamics of the above two cases, in the presence of dissipation. In both cases, the average phonon number, while still oscillatory, tends toward the open boundary at $n=0$. For the second scenario in particular, the boundary ($n=0$ unit cell) becomes significantly populated in a periodic fashion.

\section{\label{utilities}Potential utilities of NHSE}
In this section, we show that, given our proposed scheme, the dynamic features of the NHSE can have potential applications in cooling and sensing. For both cases, we consider the inevitable heating under typical experimental conditions. The heating process introduces decoherence into the external motion of the ion, typically giving rise to a higher average phonon number over time. Here we describe the heating process by introducing quantum jump processes, with the jump operators $\{L_{n,1} = \sqrt{\kappa}|n\rangle\langle n+1|, L_{n,2} = \sqrt{\kappa}|n+1\rangle\langle n|\}$, where $\kappa=2\pi\times0.3$~kHz is the typical heating rate. The dynamics of the system (still under post selection) is thus captured by a hybrid master equation
\begin{align}
\frac{\mathrm{d} \rho }{\mathrm{d} t} &=-i(H\rho-\rho H^{\dagger}) \nonumber\\
&+ \sum_{n,j} [L_{n,j}\rho L_{n,j}^{\dagger}\rho-\frac{1}{2}(L_{n,j}^{\dagger}L_{n,j}+ \rho L_{n,j}^{\dagger}L_{n.j})], \label{eqn:master}
\end{align}
where $\rho$ is the density matrix; $H$ is the effective non-Hermitian Hamiltonian (\ref{eq:H}).

\begin{figure}[htbp]
\includegraphics[width=8.6cm]{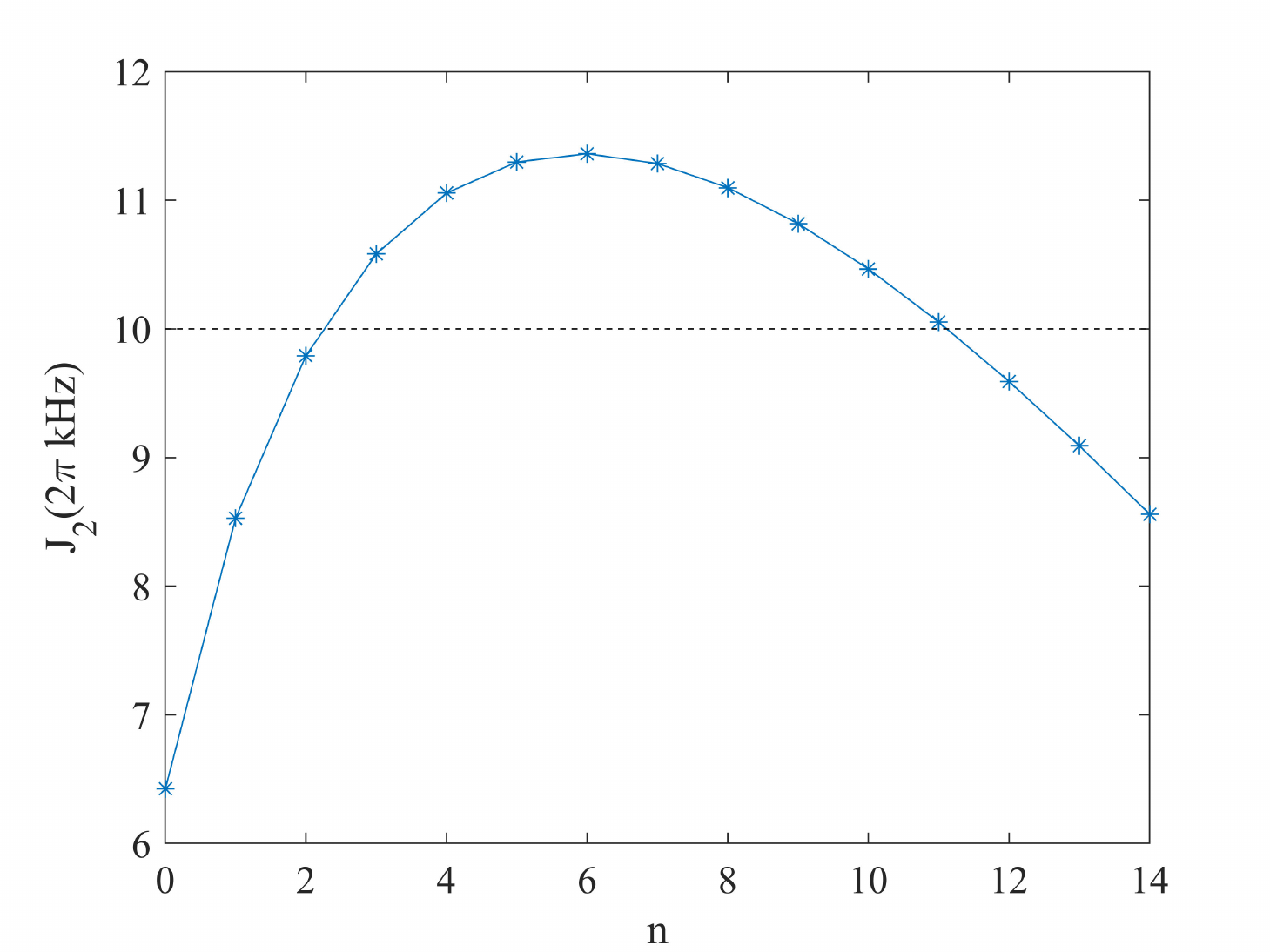}
\caption{ Inhomogeneous intercell hopping rate $J_2$ calculated through $J_2(n)=\frac{\hbar}{2}\Omega_4 \eta e^{-\eta^2/2}\sqrt{\frac{1}{n+1}} L_{n}^{1}(\eta^2)$ where $L_n^a$ is the generalized Laguerre polynomial (see Appendix). The parameters are chosen so that the average value of $J_2$ is $2\pi\times10$~kHz for $\eta=0.35$ and $n\in [0, 14]$. The horizontal dashed line indicates the mean value of $\bar{J}_2=2\pi \times 10$~kHz.}
\label{Fig_J2}
\end{figure}

Another practical consideration is the inhomogeneity in the hopping rate $J_2$. The intercell hopping realized by sideband transitions is typically inhomogeneous, which typically scales with $\sqrt{n+1}$ if the Lamb-Dicke criterion is satisfied (with the Lamb-Dicke parameter $\eta\ll 1$). As our ion of choice,  $^{9}$Be$^{+}$ gives $\eta\sim 0.35$ under typical experimental conditions (see Appendix), which does not satisfy the Lamb-Dicke criterion. In Fig.~\ref{Fig_J2}, we show the dependence of $J_2$ on $n$ in the range of $n\in [0,14]$. Apparently, $J_2$ does not scale as $\sqrt{n+1}$, and does not deviate too much from its average value in this range.
In the following, we take this inhomogeneity into account for numerical simulations.

\subsection{\label{sec:cool}NHSE for cooling}
In our setup, the NHSE-induced unidirectional flow occurs in the phonon modes. When the parameters are tuned such that the probability flows toward smaller $n$, the NHSE effectively gives rise to cooling in the external ion motion.  The question is whether the cooling effect is still significant under typical experimental conditions where heating also arises according to Eq.~(\ref{eqn:master}).

\begin{figure}[t]
\includegraphics[width=8.6cm]{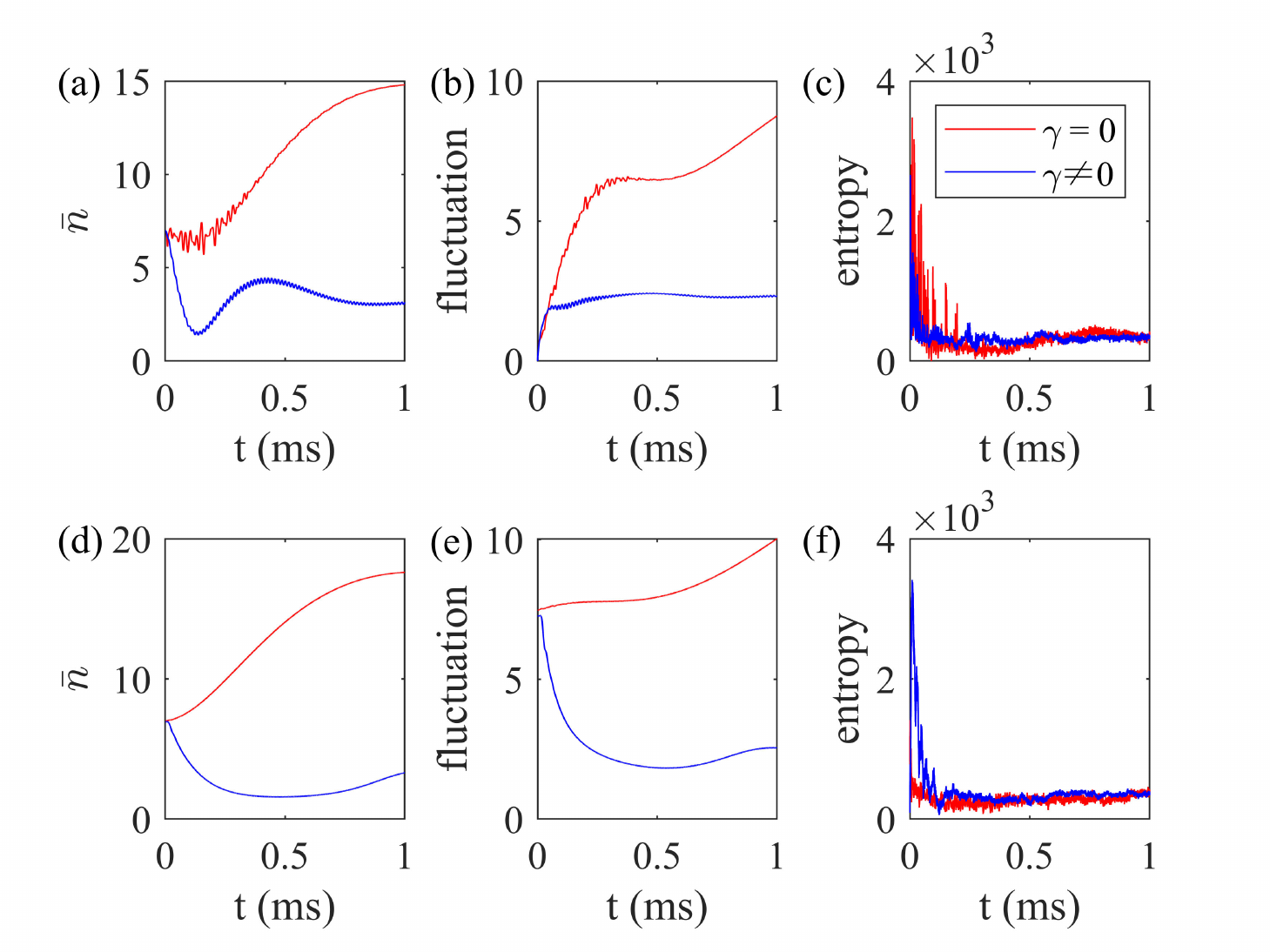}%
\caption{Ion cooling by the NHSE.
Time evolution, governed by Eq.~(\ref{eqn:master}) of (a)(d) the average phonon number, (d)(e) the fluctuations of the phonon number, and (c)(f) entropy under the NHSE.
The system is initialized in the state $\ket{n=7,a}$ for (a)(b)(c), and a thermal state with $\bar{n}_{th}=7$ for (d)(e)(f).
The red solid lines are the Hermitian cases with $\gamma = 0$, and the blue solid lines are the non-Hermitian cases with $\gamma = 2\pi\times50$~kHz. Other parameters are $J_1 = 2\pi\times20$~kHz, $\phi = -\pi/2$, and $\kappa = 2\pi\times 0.3$~kHz. The value of $J_2$ is chosen according to Fig.~\ref{Fig_J2}.}
\label{Fig5}
\end{figure}

In Fig.~\ref{Fig5}, we show the time evolution of the average phonon number, its fluctuation, and the entropy of the system, driven by the master equation Eq.~(\ref{eqn:master}). Regardless of the initial state, the average phonon number and its fluctuation oscillate slightly and approach small finite values under the impact of the NHSE. By contrast, in the absence of non-Hermiticity, both the average phonon number and the fluctuation increase during the time evolution.
From the entropy evolution, we see that in both cases, the system has already approached a quasi-steady state after $1$ ms of time evolution. Thus, the NHSE, by inducing a directional flow toward smaller $n$, effectively cools the external motion of the ion, which, while reminiscent of the resolved sideband cooling in trapped ions, derives from a distinct mechanism.

\subsection{\label{sec:sense}Sensing the synthetic magnetic flux}
While the NHSE in the dissipative AB chain originates from the interplay of dissipation and the synthetic flux,
the NHSE-induced dynamics depends sensitively on the synthetic magnetic flux, which offers the interesting possibility of quantum sensing.

\begin{figure}[t]
\includegraphics[width=8.6cm]{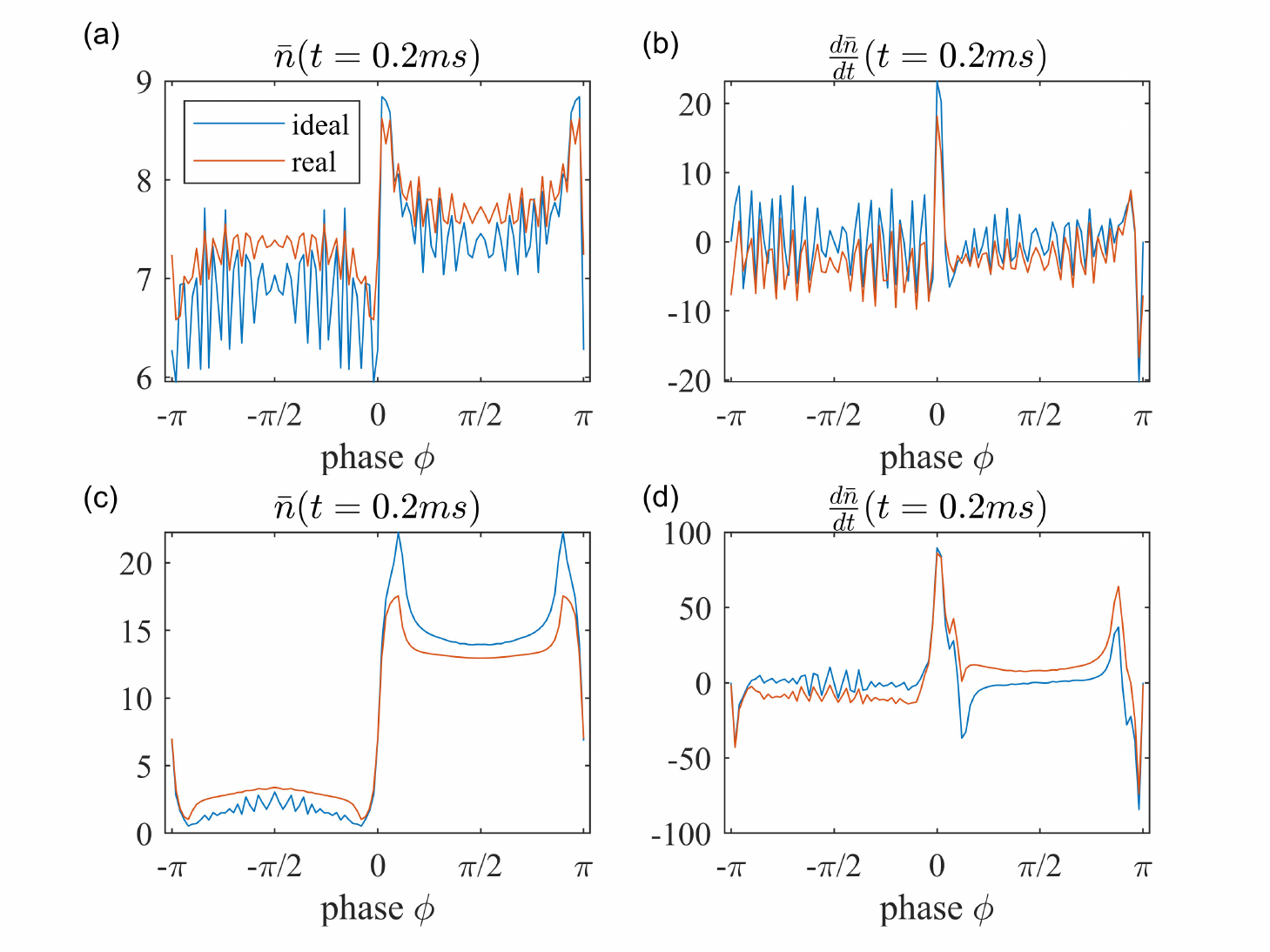}
\caption{Dependence of the NHSE on the synthetic flux.
(a)(c) The final average phonon number $\bar{n}$ as a function of the synthetic flux $\phi$, after a time evolution of $0.2$ ms.
The system is initialized in the state $\ket{n=7,a}$.
(b)(d) Derivatives of the average phonon number versus the flux.
We show the Hermitian cases in (a)(b), and take $\gamma=2\pi\times50$~kHz in (c)(d).
The ideal cases (blue curves) are the results from the theoretical dissipative AB chain model, while the real cases (red curves) include the inevitable heating and inhomogeneous intercell hopping rates in our experimental setting.
Other parameters are $J_1 = 2\pi\times100$~kHz, and $\gamma = 2\pi\times50$~kHz. The value of $J_2$ is chosen according to Fig.~\ref{Fig_J2}.}
\label{Fig4}
\end{figure}

In Fig.~\ref{Fig4}, we demonstrate the dependence of the directional flow on the flux. In particular, initializing the system in the state $\ket{n=7,a}$, we see that the average phonon number changes sharply near $\phi=0$ and $\phi=\pm \pi$. At exactly these locations, the NHSE-induces directional flow not only changes its direction, but undergoes a rapid increase/decrease in the amplitude. As such, using the average phonon number or its time derivative as the signal, one should in principle realize a sensing scheme for the synthetic magnetic flux near $\phi=0$ and $\phi=\pm \pi$. We also note that the sensitivity of the protocol becomes better with increasing $J_1/J_2$.

\section{Experimental imperfections}
For the calculations above, we have neglected the spontaneous decay back into the state $|c\rangle$.
To account for the impact of the spontaneous decay, we consider the full master equation
\begin{align}
\frac{\mathrm{d} \rho }{\mathrm{d} t} &=-i[H_{coh}, \rho] \nonumber\\
&+ \sum_{n,j} [L_{n,j}\rho L_{n,j}^{\dagger}\rho-\frac{1}{2}(L_{n,j}^{\dagger}L_{n,j}+ \rho L_{n,j}^{\dagger}L_{n.j})], \label{eqn:full master}
\end{align}
where the Hermitian Hamiltonian is
\begin{align}
H_{coh} &= \sum_n[J_1 (e^{i\phi}|n,c \rangle\langle n,a|+|n,c \rangle\langle n,b|+|n,a \rangle\langle n,b|)\nonumber\\
&+ J_2|n+1,a\rangle\langle n,b| + J_e|n,c\rangle\langle n,e|+H.c. ]
\end{align}
and the jump operators are
\begin{align}
L_{n,1} &= \sqrt{\kappa}|n\rangle\langle n+1|, L_{n,2} = \sqrt{\kappa}|n+1\rangle\langle n|,\\
L_{n,3} &= \sqrt{\Gamma_c}|c\rangle\langle e|, \\
L_{n,4} &= \sqrt{\Gamma_{r_1}}|r_1\rangle\langle e|, L_{n,5} = \sqrt{\Gamma_{r_2}}|r_2\rangle\langle e|.
\end{align}
Here $|e\rangle$, $|r_1\rangle$, $|r_2\rangle$ respectively represent the excited state $^{2}P_{1/2}\ket{F=2,m_F=2}$, and the reservoir states $\ket{r_1}$ ($\ket{F=2,m_F=2}$) and $\ket{r_2}$ ($\ket{F=2,m_F=1}$) in the $^{2}S_{1/2}$ manifold, with $\Gamma_{r_1}=\Gamma/3$, $\Gamma_{r_2}=\Gamma/6$, and $\Gamma_c=\Gamma/2$ according to the Clebsch-Gordon coefficients. The coupling rate between $\ket{c}$ and $\ket{e}$ is labeled as $J_e$, which modulates the laser-induced loss on site $c$ with an effective dissipative rate $\gamma=2J_e^2 (\Gamma_{r_1}+\Gamma_{r_2})/(\Gamma_c+\Gamma_{r_1}+\Gamma_{r_2})^2=J_e^2 /\Gamma$.

\begin{figure}[htbp]
\centering
\includegraphics[width=8.6cm]{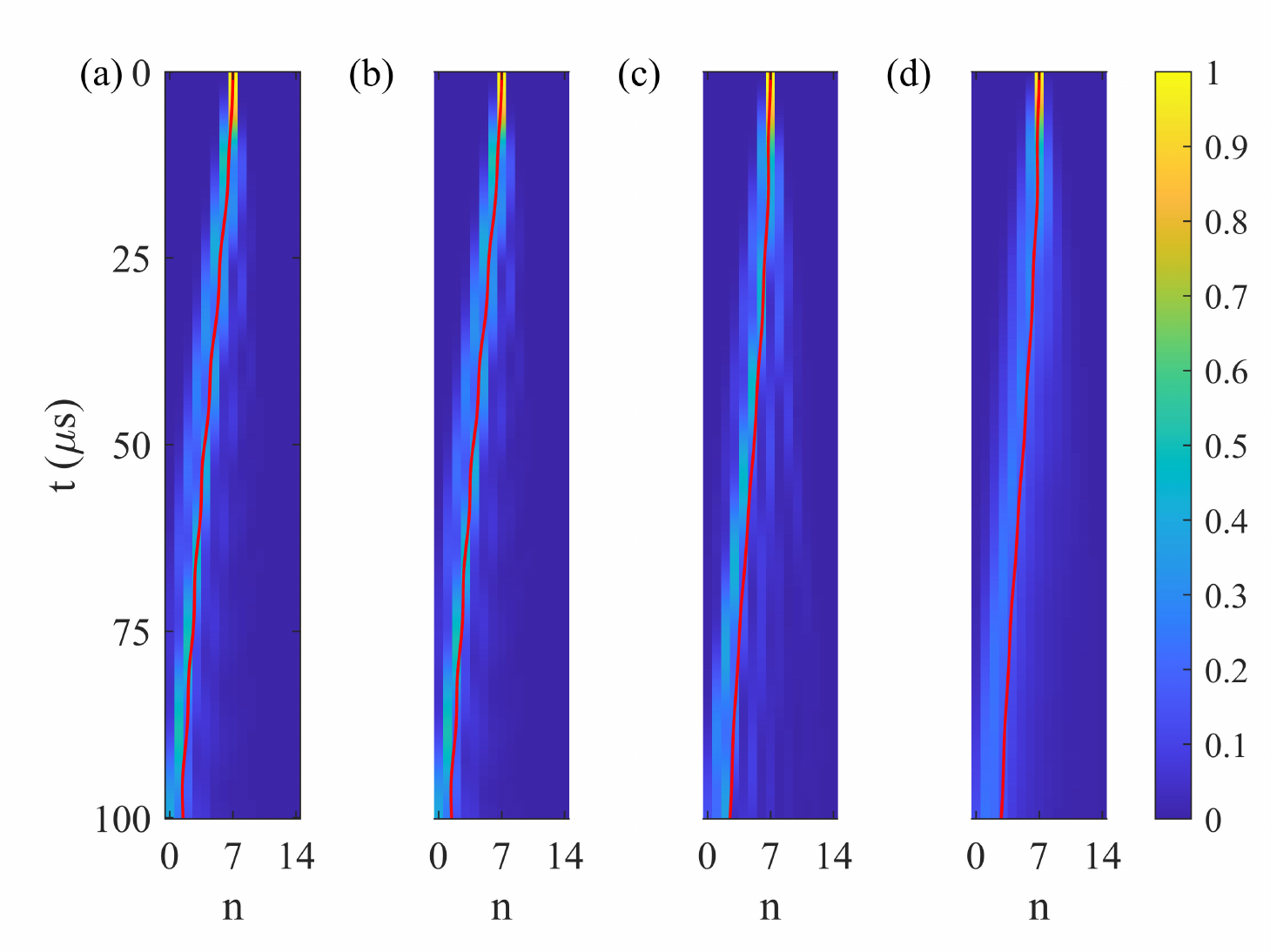}
\caption{(a) Dynamics under the non-Hermitian effective Hamiltonian~\ref{eq:H}.
(b)(c)(d) Dynamics under the Lindblad master equation Eq~(\ref{eqn:full master}), with jump operators $\{L_{n,4},L_{n,5}\}$ for (b), $\{L_{n,3},L_{n,4},L_{n,5}\}$ for (c), and $\{L_{n,1},L_{n,2},L_{n,3},L_{n,4},L_{n,5}\}$ for (d). Other parameters are $J_1 = 2\pi\times20$~kHz, $\phi = -\pi/2$,  $J_e=\sqrt{\gamma\Gamma}\approx2\pi\times0.98$~MHz, and $\gamma = 2\pi\times50$~kHz. Note that we take $\Gamma_{r_1}+\Gamma_{r_2}=\Gamma$ and $\Gamma_c=0$ in (b), such that the spontaneous decay back to the state $|c\rangle$ is neglected. The value of $J_2$ is chosen according to Fig.~\ref{Fig_J2}.
}
\label{Fig7}
\end{figure}

We now take the above imperfections into account for numerical calculations.
Figure \ref{Fig7} shows the dynamics under the non-Hermitian effective Hamiltonian~\ref{eq:H}, as illustrated in Fig.~\ref{Fig7}(a), and the Lindblad master equation Eq.~(\ref{eqn:full master}) with different sets of jump operators, as shown in Fig.~\ref{Fig7}(b)(c)(d).
For all calculations here, we adopt the inhomogeneous hopping $J_2$.
In Fig.~\ref{Fig7}(b), we only include decay from the excited state to the reservoir states, and the results are the same as those in Fig.~\ref{Fig7}(a).
Such a connection between the non-Hermitian Hamiltonian and the master equation is due to the linearity of the jump operators, as discussed previously~\cite{tianyuprb}.
In Fig.~\ref{Fig7}(c), we further introduce the decay back into the state $c$, while leaving out the jump operators for heating.
Compared with the results in Fig.~\ref{Fig7}(a)(b), dynamics in Fig.~\ref{Fig7}(c) remains largely similar. Importantly, the directional flow induced by the NHSE is still visible.
In Fig.~\ref{Fig7}(d), we include all jump operators, including those responsible for the heating processes. Again, the dynamic signatures of the NHSE is robust when both processes are present.

\begin{figure}[htbp]
\centering
\includegraphics[width=8.6cm]{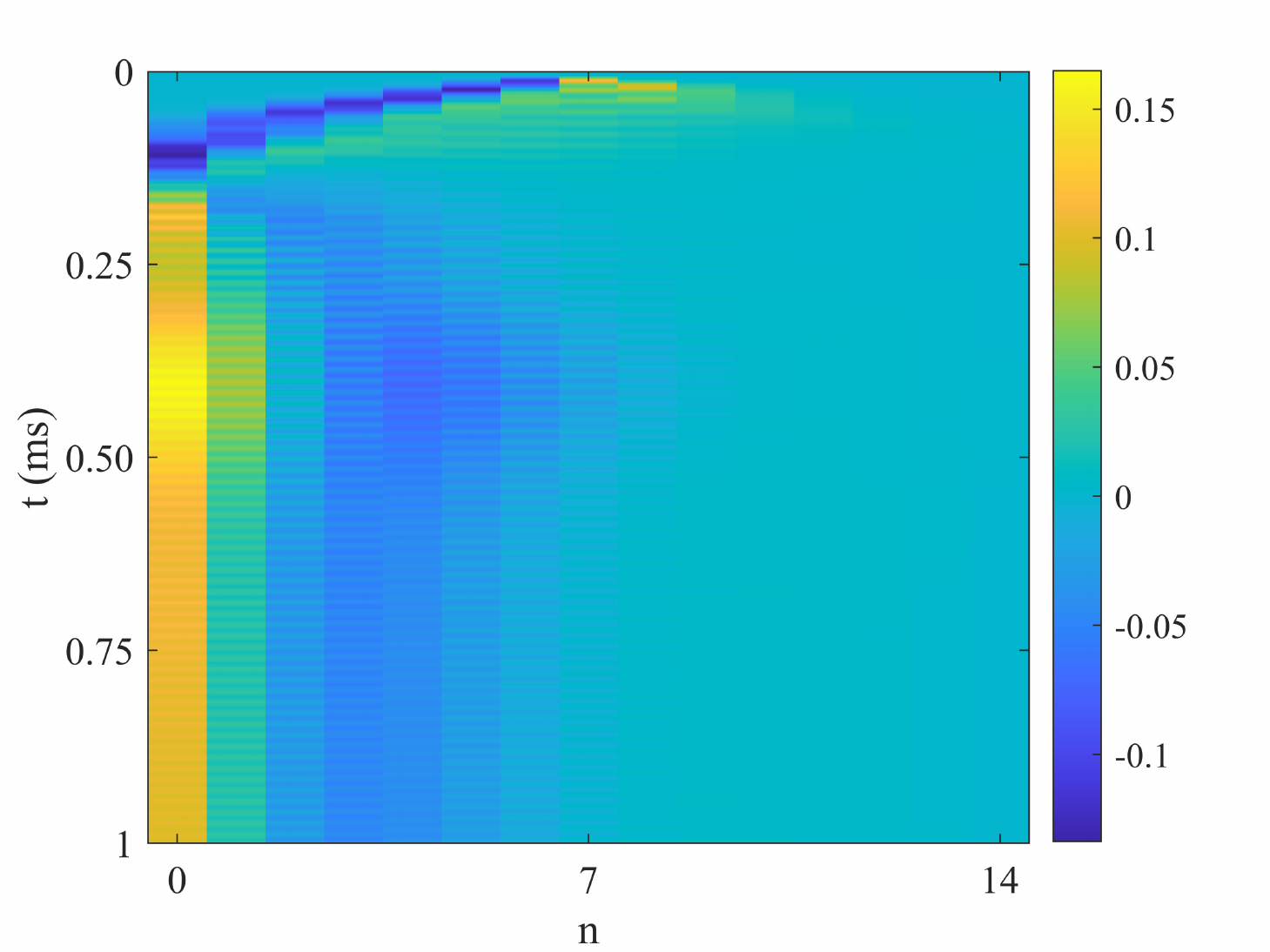}
\caption{ Error estimate from the experimental imperfection. The difference of dynamical results between the hybrid master equation~(\ref{eqn:master}) with jump operators $\{L_{n,1},L_{n,2}\}$ and the full Lindblad master equation (Fig. \ref{Fig7}(d)) is shown in population. The color bar indicates the difference of normalized probability in each cell. Other parameters are $J_1 = 2\pi\times20$~kHz, $\phi = -\pi/2$ and effectively $\gamma = 2\pi\times50$~kHz through $J_e=\sqrt{\gamma\Gamma}\approx2\pi\times0.98$~MHz. The value of $J_2$ is chosen according to Fig.~\ref{Fig_J2}. }
\label{Fig8}
\end{figure}

Figure \ref{Fig8} shows the time-evolved difference in the phonon-mode occupation, between dynamics under the hybrid master equation Eq.~(\ref{eqn:master}) with jump operators $\{L_{n,1},L_{n,2}\}$, and that under the full Lindblad master equation Eq.~(\ref{eqn:full master}), with jump operators $\{L_{n,1},L_{n,2},L_{n,3},L_{n,4},L_{n,5}\}$. At all times, the difference is much smaller than one phonon. This confirms that the experimental imperfections do not qualitatively alter the dynamic signature of the NHSE.

\section{\label{outlook}Discussions and outlook}
To summarize, we propose a practical scheme to simulate the dissipative AB chain using a single trapped ion. We focus on the NHSE of the model, and demonstrate that dynamic signatures of NHSE can be detected by probing the occupation of the phonon modes. By analyzing in detail the dependence of the directional flow on system parameters, we show that the flow can in principle be used for cooling and sensing.

For future studies, it would be interesting to exploit the band topology of the AB chain for quantum simulation and manipulation. Since the system of trapped ions is famed for its scalability, our proposal also paves the way for implementing synthetic non-Hermitian models in higher dimensions~\cite{doi.org/10.1038/s41467-022-30161-6}, and with few- or many-body correlations. This can be achieved, for instance, by realizing a chain of trapped ions, each individually engineered to simulate a dissipative AB chain. The extra degrees of freedom and the Coulomb interaction between ions~\cite{RevModPhys.93.025001}, while both highly tunable, offer a versatile platform for theoretical and experimental study.

\begin{acknowledgments}
We acknowledge support from the National Natural Science Foundation of China (grants No. 92165206, No. 11974330, No. 11974331), the Chinese Academy of Sciences (Grant No. XDC07000000), Innovation Program for Quantum Science and Technology (Grant No. 2021ZD0301603), Anhui Initiative in Quantum Information Technologies (Grant No. AHY050000), the Fundamental Research Funds for the Central Universities, and Hefei Comprehensive National Science Center. W.Y. acknowledges support from the National Key R\&D Program (Grant Nos. 2017YFA0304100).
\end{acknowledgments}

\appendix

\begin{widetext}
\section{Derivation of the effective Hamiltonian}
We start by writing down the Hamiltonian of the level scheme in Fig.~\ref{Fig1}, where under an appropriate rotating frame

\begin{equation*}
\begin{aligned}
H_1 =
\frac{\hbar}{2}(\Omega_1 |a \rangle\langle b|
+\Omega_2 e^{i\phi}|c \rangle\langle a|
+\Omega_3 |c \rangle\langle b|
+\Omega_4 e^{i(\Delta k_z z-\frac{\pi}{2})} e^{-i\tilde{\delta} t} |a \rangle\langle b| )+H.c.
\end{aligned}.
\end{equation*}
Here we assume that the effective coupling rates $\Omega_i$ ($i=1,2,3,4$) are real, and $\Delta k_z$ is the wave-vector difference, along the $z$-axis, between the Raman lasers (red in Fig.~\ref{Fig1}). $\tilde{\delta}$ is the detuning in the Raman two-photon transition.

Now we consider ion's harmonic motion along the axis $z$, whereas its motion in the $x$-$y$ plane is considered to be frozen. We have
\begin{equation*}
\begin{aligned}
H_2 =& \sum_n n\hbar \omega_z (|n,a \rangle\langle n,a|+|n,b \rangle\langle n,b| +|n,c \rangle\langle n,c| )
+ \sum_n \frac{\hbar}{2} (\Omega_1|n,a \rangle\langle n,b|+\Omega_2 e^{i\phi}|n,c \rangle\langle n,a| \\
&+\Omega_3 |n,c \rangle\langle n,b|+H.c. )
+\sum_{m,n} \frac{\hbar}{2} (\Omega_4  \langle m|e^{i\eta(a^{\dag}+a)}|n\rangle e^{-i\tilde{\delta} t-i\frac{\pi}{2}}|m,a\rangle\langle n,b|+H.c.),
\end{aligned}
\end{equation*}
where $\omega_z$ is the trapping frequency, and the Lamb-Dicke parameter $\eta$ is defined through
$\Delta k_z z=\Delta k_z\sqrt{\frac{1}{2m\omega_z}}(a^{\dag}+a)=\eta (a^{\dag}+a)$.

Importantly, the intercell hopping rate is calculated through
\begin{equation*}
\begin{aligned}
\langle m|e^{i\eta(a^{\dag}+a)}|n\rangle&=\langle m|D(i\eta)|n\rangle=(i\eta)^{|m-n|} e^{-\eta^2/2}\sqrt{\frac{n_<!}{n_>!}} L_{n_<}^{|m-n|}(\eta^2),
\end{aligned}
\end{equation*}
where $n_<(n_>)$ is the lesser (greater) of $m$ and $n$, and $L_n^a$ is the generalized Laguerre polynomial.

Performing the unitary transformation $U=\exp(i\tilde{\delta} a^{\dag}at)$, and taking the rotating-wave approximation, we have
\begin{align}
H_3 &= UH_3 U^{\dag} -iU\frac{dU^{\dag}}{dt}\\
&=\sum_n n\delta(|n,a \rangle\langle n,a|+|n,b \rangle\langle n,b| +|n,c \rangle\langle n,c| )+
\sum_n \frac{\hbar}{2} (\Omega_1|n,a \rangle\langle n,b|+\Omega_2 e^{i\phi}|n,c \rangle\langle n,a| +\Omega_3 |n,c \rangle\langle n,b|\nonumber\\
&+ 0.48\Omega_4 |n+1,a\rangle\langle n,b|)+H.c.\\
&\equiv\sum_n[J_1 (e^{i\phi}|n,c \rangle\langle n,a|+|n,c \rangle\langle n,b|+|n,a \rangle\langle n,b|)+ J_2|n+1,a\rangle\langle n,b|+H.c. ],
\end{align}
where $\delta=\omega_z-\tilde{\delta}$. Here we have taken $\eta=0.35$, typical for a trapped $^{9}$Be$^{+}$ ion. Since $\langle n+1|e^{i\eta(a^{\dag}+a)}|n\rangle e^{-i\frac{\pi}{2}}=\eta e^{-\eta^2/2}\sqrt{\frac{1}{n+1}} L_{n}^{1}(\eta^2)\approx 0.48$ for $\eta=0.35$ and $n\in [0, 14]$, we neglect the heterogeneity in the intercell coupling rate.

For the laser-induced dissipation, the spontaneous decay from the excited state $\ket{F=2,m_F=2}$ ($^{2}P_{1/2}$) ends up in the states $\ket{c}$, $\ket{r_1}$ ($\ket{F=2,m_F=2}$) and $\ket{r_2}$ ($\ket{F=2,m_F=1}$) in the ground state manifold, with the branching ratio $3:2:1$. Denoting the linewidth of the excited state as $\Gamma$, the effective laser-induced loss rate is then $\gamma=J_e^2 /\Gamma$, where $J_e$ is the coupling strength of the pumping laser.

Finally, we have the non-Hermitian effective Hamiltonian that is the tilted AB chain model
\begin{align}
H &=\sum_n n\delta(|n,a \rangle\langle n,a|+|n,b \rangle\langle n,b| +|n,c \rangle\langle n,c| ) -\sum_n i\gamma |n,c \rangle\langle n,c|\nonumber \\
&+ \sum_n[J_1 (e^{i\phi}|n,c \rangle\langle n,a|+|n,c \rangle\langle n,b|+|n,a \rangle\langle n,b|)+ J_2|n+1,a\rangle\langle n,b|+H.c. ].
\end{align}
\end{widetext}


\begin{thebibliography}{36}%
\makeatletter
\providecommand \@ifxundefined [1]{%
 \@ifx{#1\undefined}
}%
\providecommand \@ifnum [1]{%
 \ifnum #1\expandafter \@firstoftwo
 \else \expandafter \@secondoftwo
 \fi
}%
\providecommand \@ifx [1]{%
 \ifx #1\expandafter \@firstoftwo
 \else \expandafter \@secondoftwo
 \fi
}%
\providecommand \natexlab [1]{#1}%
\providecommand \enquote  [1]{``#1''}%
\providecommand \bibnamefont  [1]{#1}%
\providecommand \bibfnamefont [1]{#1}%
\providecommand \citenamefont [1]{#1}%
\providecommand \href@noop [0]{\@secondoftwo}%
\providecommand \href [0]{\begingroup \@sanitize@url \@href}%
\providecommand \@href[1]{\@@startlink{#1}\@@href}%
\providecommand \@@href[1]{\endgroup#1\@@endlink}%
\providecommand \@sanitize@url [0]{\catcode `\\12\catcode `\$12\catcode
  `\&12\catcode `\#12\catcode `\^12\catcode `\_12\catcode `\%12\relax}%
\providecommand \@@startlink[1]{}%
\providecommand \@@endlink[0]{}%
\providecommand \url  [0]{\begingroup\@sanitize@url \@url }%
\providecommand \@url [1]{\endgroup\@href {#1}{\urlprefix }}%
\providecommand \urlprefix  [0]{URL }%
\providecommand \Eprint [0]{\href }%
\providecommand \doibase [0]{https://doi.org/}%
\providecommand \selectlanguage [0]{\@gobble}%
\providecommand \bibinfo  [0]{\@secondoftwo}%
\providecommand \bibfield  [0]{\@secondoftwo}%
\providecommand \translation [1]{[#1]}%
\providecommand \BibitemOpen [0]{}%
\providecommand \bibitemStop [0]{}%
\providecommand \bibitemNoStop [0]{.\EOS\space}%
\providecommand \EOS [0]{\spacefactor3000\relax}%
\providecommand \BibitemShut  [1]{\csname bibitem#1\endcsname}%
\let\auto@bib@innerbib\@empty
\bibitem [{\citenamefont {Yao}\ and\ \citenamefont
  {Wang}(2018)}]{PhysRevLett.121.086803}%
  \BibitemOpen
  \bibfield  {author} {\bibinfo {author} {\bibfnamefont {S.}~\bibnamefont
  {Yao}}\ and\ \bibinfo {author} {\bibfnamefont {Z.}~\bibnamefont {Wang}},\
  }\bibfield  {title} {\bibinfo {title} {Edge states and topological invariants
  of non-hermitian systems},\ }\href
  {https://doi.org/10.1103/PhysRevLett.121.086803} {\bibfield  {journal}
  {\bibinfo  {journal} {Phys. Rev. Lett.}\ }\textbf {\bibinfo {volume} {121}},\
  \bibinfo {pages} {086803} (\bibinfo {year} {2018})}\BibitemShut {NoStop}%
\bibitem [{\citenamefont {Yokomizo}\ and\ \citenamefont
  {Murakami}(2019)}]{PhysRevLett.123.066404}%
  \BibitemOpen
  \bibfield  {author} {\bibinfo {author} {\bibfnamefont {K.}~\bibnamefont
  {Yokomizo}}\ and\ \bibinfo {author} {\bibfnamefont {S.}~\bibnamefont
  {Murakami}},\ }\bibfield  {title} {\bibinfo {title} {Non-bloch band theory of
  non-hermitian systems},\ }\href
  {https://doi.org/10.1103/PhysRevLett.123.066404} {\bibfield  {journal}
  {\bibinfo  {journal} {Phys. Rev. Lett.}\ }\textbf {\bibinfo {volume} {123}},\
  \bibinfo {pages} {066404} (\bibinfo {year} {2019})}\BibitemShut {NoStop}%
\bibitem [{\citenamefont {Yao}\ \emph {et~al.}(2018)\citenamefont {Yao},
  \citenamefont {Song},\ and\ \citenamefont {Wang}}]{PhysRevLett.121.136802}%
  \BibitemOpen
  \bibfield  {author} {\bibinfo {author} {\bibfnamefont {S.}~\bibnamefont
  {Yao}}, \bibinfo {author} {\bibfnamefont {F.}~\bibnamefont {Song}},\ and\
  \bibinfo {author} {\bibfnamefont {Z.}~\bibnamefont {Wang}},\ }\bibfield
  {title} {\bibinfo {title} {Non-hermitian chern bands},\ }\href
  {https://doi.org/10.1103/PhysRevLett.121.136802} {\bibfield  {journal}
  {\bibinfo  {journal} {Phys. Rev. Lett.}\ }\textbf {\bibinfo {volume} {121}},\
  \bibinfo {pages} {136802} (\bibinfo {year} {2018})}\BibitemShut {NoStop}%
\bibitem [{\citenamefont {Lee}\ and\ \citenamefont
  {Thomale}(2019)}]{PhysRevB.99.201103}%
  \BibitemOpen
  \bibfield  {author} {\bibinfo {author} {\bibfnamefont {C.~H.}\ \bibnamefont
  {Lee}}\ and\ \bibinfo {author} {\bibfnamefont {R.}~\bibnamefont {Thomale}},\
  }\bibfield  {title} {\bibinfo {title} {Anatomy of skin modes and topology in
  non-hermitian systems},\ }\href {https://doi.org/10.1103/PhysRevB.99.201103}
  {\bibfield  {journal} {\bibinfo  {journal} {Phys. Rev. B}\ }\textbf {\bibinfo
  {volume} {99}},\ \bibinfo {pages} {201103(R)} (\bibinfo {year}
  {2019})}\BibitemShut {NoStop}%
\bibitem [{\citenamefont {Kunst}\ \emph {et~al.}(2018)\citenamefont {Kunst},
  \citenamefont {Edvardsson}, \citenamefont {Budich},\ and\ \citenamefont
  {Bergholtz}}]{PhysRevLett.121.026808}%
  \BibitemOpen
  \bibfield  {author} {\bibinfo {author} {\bibfnamefont {F.~K.}\ \bibnamefont
  {Kunst}}, \bibinfo {author} {\bibfnamefont {E.}~\bibnamefont {Edvardsson}},
  \bibinfo {author} {\bibfnamefont {J.~C.}\ \bibnamefont {Budich}},\ and\
  \bibinfo {author} {\bibfnamefont {E.~J.}\ \bibnamefont {Bergholtz}},\
  }\bibfield  {title} {\bibinfo {title} {Biorthogonal bulk-boundary
  correspondence in non-hermitian systems},\ }\href
  {https://doi.org/10.1103/PhysRevLett.121.026808} {\bibfield  {journal}
  {\bibinfo  {journal} {Phys. Rev. Lett.}\ }\textbf {\bibinfo {volume} {121}},\
  \bibinfo {pages} {026808} (\bibinfo {year} {2018})}\BibitemShut {NoStop}%
\bibitem [{\citenamefont {McDonald}\ \emph {et~al.}(2018)\citenamefont
  {McDonald}, \citenamefont {Pereg-Barnea},\ and\ \citenamefont
  {Clerk}}]{PhysRevX.8.041031}%
  \BibitemOpen
  \bibfield  {author} {\bibinfo {author} {\bibfnamefont {A.}~\bibnamefont
  {McDonald}}, \bibinfo {author} {\bibfnamefont {T.}~\bibnamefont
  {Pereg-Barnea}},\ and\ \bibinfo {author} {\bibfnamefont {A.~A.}\ \bibnamefont
  {Clerk}},\ }\bibfield  {title} {\bibinfo {title} {Phase-dependent chiral
  transport and effective non-hermitian dynamics in a bosonic kitaev-majorana
  chain},\ }\href {https://doi.org/10.1103/PhysRevX.8.041031} {\bibfield
  {journal} {\bibinfo  {journal} {Phys. Rev. X}\ }\textbf {\bibinfo {volume}
  {8}},\ \bibinfo {pages} {041031} (\bibinfo {year} {2018})}\BibitemShut
  {NoStop}%
\bibitem [{\citenamefont {Martinez~Alvarez}\ \emph {et~al.}(2018)\citenamefont
  {Martinez~Alvarez}, \citenamefont {Barrios~Vargas},\ and\ \citenamefont
  {Foa~Torres}}]{PhysRevB.97.121401}%
  \BibitemOpen
  \bibfield  {author} {\bibinfo {author} {\bibfnamefont {V.~M.}\ \bibnamefont
  {Martinez~Alvarez}}, \bibinfo {author} {\bibfnamefont {J.~E.}\ \bibnamefont
  {Barrios~Vargas}},\ and\ \bibinfo {author} {\bibfnamefont {L.~E.~F.}\
  \bibnamefont {Foa~Torres}},\ }\bibfield  {title} {\bibinfo {title}
  {Non-hermitian robust edge states in one dimension: Anomalous localization
  and eigenspace condensation at exceptional points},\ }\href
  {https://doi.org/10.1103/PhysRevB.97.121401} {\bibfield  {journal} {\bibinfo
  {journal} {Phys. Rev. B}\ }\textbf {\bibinfo {volume} {97}},\ \bibinfo
  {pages} {121401(R)} (\bibinfo {year} {2018})}\BibitemShut {NoStop}%
\bibitem [{\citenamefont {Zhang}\ \emph {et~al.}(2020)\citenamefont {Zhang},
  \citenamefont {Yang},\ and\ \citenamefont {Fang}}]{PhysRevLett.125.126402}%
  \BibitemOpen
  \bibfield  {author} {\bibinfo {author} {\bibfnamefont {K.}~\bibnamefont
  {Zhang}}, \bibinfo {author} {\bibfnamefont {Z.}~\bibnamefont {Yang}},\ and\
  \bibinfo {author} {\bibfnamefont {C.}~\bibnamefont {Fang}},\ }\bibfield
  {title} {\bibinfo {title} {Correspondence between winding numbers and skin
  modes in non-hermitian systems},\ }\href
  {https://doi.org/10.1103/PhysRevLett.125.126402} {\bibfield  {journal}
  {\bibinfo  {journal} {Phys. Rev. Lett.}\ }\textbf {\bibinfo {volume} {125}},\
  \bibinfo {pages} {126402} (\bibinfo {year} {2020})}\BibitemShut {NoStop}%
\bibitem [{\citenamefont {Okuma}\ \emph {et~al.}(2020)\citenamefont {Okuma},
  \citenamefont {Kawabata}, \citenamefont {Shiozaki},\ and\ \citenamefont
  {Sato}}]{PhysRevLett.124.086801}%
  \BibitemOpen
  \bibfield  {author} {\bibinfo {author} {\bibfnamefont {N.}~\bibnamefont
  {Okuma}}, \bibinfo {author} {\bibfnamefont {K.}~\bibnamefont {Kawabata}},
  \bibinfo {author} {\bibfnamefont {K.}~\bibnamefont {Shiozaki}},\ and\
  \bibinfo {author} {\bibfnamefont {M.}~\bibnamefont {Sato}},\ }\bibfield
  {title} {\bibinfo {title} {Topological origin of non-hermitian skin
  effects},\ }\href {https://doi.org/10.1103/PhysRevLett.124.086801} {\bibfield
   {journal} {\bibinfo  {journal} {Phys. Rev. Lett.}\ }\textbf {\bibinfo
  {volume} {124}},\ \bibinfo {pages} {086801} (\bibinfo {year}
  {2020})}\BibitemShut {NoStop}%
\bibitem [{\citenamefont {Longhi}(2019)}]{PhysRevResearch.1.023013}%
  \BibitemOpen
  \bibfield  {author} {\bibinfo {author} {\bibfnamefont {S.}~\bibnamefont
  {Longhi}},\ }\bibfield  {title} {\bibinfo {title} {Probing non-hermitian skin
  effect and non-bloch phase transitions},\ }\href
  {https://doi.org/10.1103/PhysRevResearch.1.023013} {\bibfield  {journal}
  {\bibinfo  {journal} {Phys. Rev. Research}\ }\textbf {\bibinfo {volume}
  {1}},\ \bibinfo {pages} {023013} (\bibinfo {year} {2019})}\BibitemShut
  {NoStop}%
\bibitem [{\citenamefont {Yang}\ \emph {et~al.}(2020)\citenamefont {Yang},
  \citenamefont {Zhang}, \citenamefont {Fang},\ and\ \citenamefont
  {Hu}}]{PhysRevLett.125.226402}%
  \BibitemOpen
  \bibfield  {author} {\bibinfo {author} {\bibfnamefont {Z.}~\bibnamefont
  {Yang}}, \bibinfo {author} {\bibfnamefont {K.}~\bibnamefont {Zhang}},
  \bibinfo {author} {\bibfnamefont {C.}~\bibnamefont {Fang}},\ and\ \bibinfo
  {author} {\bibfnamefont {J.}~\bibnamefont {Hu}},\ }\bibfield  {title}
  {\bibinfo {title} {Non-hermitian bulk-boundary correspondence and auxiliary
  generalized brillouin zone theory},\ }\href
  {https://doi.org/10.1103/PhysRevLett.125.226402} {\bibfield  {journal}
  {\bibinfo  {journal} {Phys. Rev. Lett.}\ }\textbf {\bibinfo {volume} {125}},\
  \bibinfo {pages} {226402} (\bibinfo {year} {2020})}\BibitemShut {NoStop}%
\bibitem [{\citenamefont {Deng}\ and\ \citenamefont
  {Yi}(2019)}]{PhysRevB.100.035102}%
  \BibitemOpen
  \bibfield  {author} {\bibinfo {author} {\bibfnamefont {T.-S.}\ \bibnamefont
  {Deng}}\ and\ \bibinfo {author} {\bibfnamefont {W.}~\bibnamefont {Yi}},\
  }\bibfield  {title} {\bibinfo {title} {Non-bloch topological invariants in a
  non-hermitian domain wall system},\ }\href
  {https://doi.org/10.1103/PhysRevB.100.035102} {\bibfield  {journal} {\bibinfo
   {journal} {Phys. Rev. B}\ }\textbf {\bibinfo {volume} {100}},\ \bibinfo
  {pages} {035102} (\bibinfo {year} {2019})}\BibitemShut {NoStop}%
\bibitem [{\citenamefont {Li}\ \emph {et~al.}(2020)\citenamefont {Li},
  \citenamefont {Lee}, \citenamefont {Mu},\ and\ \citenamefont
  {Gong}}]{Nature.communications.11.1.(2020)}%
  \BibitemOpen
  \bibfield  {author} {\bibinfo {author} {\bibfnamefont {L.}~\bibnamefont
  {Li}}, \bibinfo {author} {\bibfnamefont {C.~H.}\ \bibnamefont {Lee}},
  \bibinfo {author} {\bibfnamefont {S.}~\bibnamefont {Mu}},\ and\ \bibinfo
  {author} {\bibfnamefont {J.}~\bibnamefont {Gong}},\ }\bibfield  {title}
  {\bibinfo {title} {Critical non-hermitian skin effect},\ }\href
  {https://doi.org/10.1038/s41467-020-18917-4} {\bibfield  {journal} {\bibinfo
  {journal} {Nat. Commun.}\ }\textbf {\bibinfo {volume} {11}},\ \bibinfo
  {pages} {5491} (\bibinfo {year} {2020})}\BibitemShut {NoStop}%
\bibitem [{\citenamefont {Zhou}\ \emph {et~al.}(2021)\citenamefont {Zhou},
  \citenamefont {Li}, \citenamefont {Yi},\ and\ \citenamefont
  {Cui}}]{2111.04196}%
  \BibitemOpen
  \bibfield  {author} {\bibinfo {author} {\bibfnamefont {L.}~\bibnamefont
  {Zhou}}, \bibinfo {author} {\bibfnamefont {H.}~\bibnamefont {Li}}, \bibinfo
  {author} {\bibfnamefont {W.}~\bibnamefont {Yi}},\ and\ \bibinfo {author}
  {\bibfnamefont {X.}~\bibnamefont {Cui}},\ }\href@noop {} {\bibinfo {title}
  {Engineering non-hermitian skin effect with band topology in ultracold
  gases}} (\bibinfo {year} {2021}),\ \Eprint
  {https://arxiv.org/abs/arXiv:2111.04196} {arXiv:2111.04196} \BibitemShut
  {NoStop}%
\bibitem [{\citenamefont {Guo}\ \emph {et~al.}(2021)\citenamefont {Guo},
  \citenamefont {Dong}, \citenamefont {Zhang}, \citenamefont {Hu},\ and\
  \citenamefont {Yang}}]{2111.04220}%
  \BibitemOpen
  \bibfield  {author} {\bibinfo {author} {\bibfnamefont {S.}~\bibnamefont
  {Guo}}, \bibinfo {author} {\bibfnamefont {C.}~\bibnamefont {Dong}}, \bibinfo
  {author} {\bibfnamefont {F.}~\bibnamefont {Zhang}}, \bibinfo {author}
  {\bibfnamefont {J.}~\bibnamefont {Hu}},\ and\ \bibinfo {author}
  {\bibfnamefont {Z.}~\bibnamefont {Yang}},\ }\href@noop {} {\bibinfo {title}
  {Theoretical prediction of non-hermitian skin effect in ultracold atom
  systems}} (\bibinfo {year} {2021}),\ \Eprint
  {https://arxiv.org/abs/arXiv:2111.04220} {arXiv:2111.04220} \BibitemShut
  {NoStop}%
\bibitem [{\citenamefont {Li}\ \emph {et~al.}(2021)\citenamefont {Li},
  \citenamefont {Sun}, \citenamefont {Zhang},\ and\ \citenamefont
  {Yi}}]{PhysRevResearch.3.023022}%
  \BibitemOpen
  \bibfield  {author} {\bibinfo {author} {\bibfnamefont {T.}~\bibnamefont
  {Li}}, \bibinfo {author} {\bibfnamefont {J.-Z.}\ \bibnamefont {Sun}},
  \bibinfo {author} {\bibfnamefont {Y.-S.}\ \bibnamefont {Zhang}},\ and\
  \bibinfo {author} {\bibfnamefont {W.}~\bibnamefont {Yi}},\ }\bibfield
  {title} {\bibinfo {title} {Non-bloch quench dynamics},\ }\href
  {https://doi.org/10.1103/PhysRevResearch.3.023022} {\bibfield  {journal}
  {\bibinfo  {journal} {Phys. Rev. Research}\ }\textbf {\bibinfo {volume}
  {3}},\ \bibinfo {pages} {023022} (\bibinfo {year} {2021})}\BibitemShut
  {NoStop}%
\bibitem [{\citenamefont {Song}\ \emph {et~al.}(2019)\citenamefont {Song},
  \citenamefont {Yao},\ and\ \citenamefont {Wang}}]{PhysRevLett.123.170401}%
  \BibitemOpen
  \bibfield  {author} {\bibinfo {author} {\bibfnamefont {F.}~\bibnamefont
  {Song}}, \bibinfo {author} {\bibfnamefont {S.}~\bibnamefont {Yao}},\ and\
  \bibinfo {author} {\bibfnamefont {Z.}~\bibnamefont {Wang}},\ }\bibfield
  {title} {\bibinfo {title} {Non-hermitian skin effect and chiral damping in
  open quantum systems},\ }\href
  {https://doi.org/10.1103/PhysRevLett.123.170401} {\bibfield  {journal}
  {\bibinfo  {journal} {Phys. Rev. Lett.}\ }\textbf {\bibinfo {volume} {123}},\
  \bibinfo {pages} {170401} (\bibinfo {year} {2019})}\BibitemShut {NoStop}%
\bibitem [{\citenamefont {Longhi}(2020{\natexlab{a}})}]{PhysRevB.102.201103}%
  \BibitemOpen
  \bibfield  {author} {\bibinfo {author} {\bibfnamefont {S.}~\bibnamefont
  {Longhi}},\ }\bibfield  {title} {\bibinfo {title} {Unraveling the
  non-hermitian skin effect in dissipative systems},\ }\href
  {https://doi.org/10.1103/PhysRevB.102.201103} {\bibfield  {journal} {\bibinfo
   {journal} {Phys. Rev. B}\ }\textbf {\bibinfo {volume} {102}},\ \bibinfo
  {pages} {201103(R)} (\bibinfo {year} {2020}{\natexlab{a}})}\BibitemShut
  {NoStop}%
\bibitem [{\citenamefont
  {Longhi}(2020{\natexlab{b}})}]{PhysRevLett.124.066602}%
  \BibitemOpen
  \bibfield  {author} {\bibinfo {author} {\bibfnamefont {S.}~\bibnamefont
  {Longhi}},\ }\bibfield  {title} {\bibinfo {title} {Non-bloch-band collapse
  and chiral zener tunneling},\ }\href
  {https://doi.org/10.1103/PhysRevLett.124.066602} {\bibfield  {journal}
  {\bibinfo  {journal} {Phys. Rev. Lett.}\ }\textbf {\bibinfo {volume} {124}},\
  \bibinfo {pages} {066602} (\bibinfo {year} {2020}{\natexlab{b}})}\BibitemShut
  {NoStop}%
\bibitem [{\citenamefont {Helbig}\ \emph {et~al.}(2020)\citenamefont {Helbig},
  \citenamefont {Hofmann}, \citenamefont {Imhof}, \citenamefont {Abdelghany},
  \citenamefont {Kiessling}, \citenamefont {Molenkamp}, \citenamefont {Lee},
  \citenamefont {Szameit}, \citenamefont {Greiter},\ and\ \citenamefont
  {Thomale}}]{Nat.Phys.16.747}%
  \BibitemOpen
  \bibfield  {author} {\bibinfo {author} {\bibfnamefont {T.}~\bibnamefont
  {Helbig}}, \bibinfo {author} {\bibfnamefont {T.}~\bibnamefont {Hofmann}},
  \bibinfo {author} {\bibfnamefont {S.}~\bibnamefont {Imhof}}, \bibinfo
  {author} {\bibfnamefont {M.}~\bibnamefont {Abdelghany}}, \bibinfo {author}
  {\bibfnamefont {T.}~\bibnamefont {Kiessling}}, \bibinfo {author}
  {\bibfnamefont {L.~W.}\ \bibnamefont {Molenkamp}}, \bibinfo {author}
  {\bibfnamefont {C.~H.}\ \bibnamefont {Lee}}, \bibinfo {author} {\bibfnamefont
  {A.}~\bibnamefont {Szameit}}, \bibinfo {author} {\bibfnamefont
  {M.}~\bibnamefont {Greiter}},\ and\ \bibinfo {author} {\bibfnamefont
  {R.}~\bibnamefont {Thomale}},\ }\bibfield  {title} {\bibinfo {title}
  {Generalized bulk–boundary correspondence in non-hermitian topolectrical
  circuits},\ }\href {https://doi.org/10.1038/s41567-020-0922-9} {\bibfield
  {journal} {\bibinfo  {journal} {Nat. Phys.}\ }\textbf {\bibinfo {volume}
  {16}},\ \bibinfo {pages} {747} (\bibinfo {year} {2020})}\BibitemShut
  {NoStop}%
\bibitem [{\citenamefont {Ghatak}\ \emph {et~al.}(2020)\citenamefont {Ghatak},
  \citenamefont {Brandenbourger}, \citenamefont {van Wezel},\ and\
  \citenamefont {Coulais}}]{doi:10.1073/pnas.2010580117}%
  \BibitemOpen
  \bibfield  {author} {\bibinfo {author} {\bibfnamefont {A.}~\bibnamefont
  {Ghatak}}, \bibinfo {author} {\bibfnamefont {M.}~\bibnamefont
  {Brandenbourger}}, \bibinfo {author} {\bibfnamefont {J.}~\bibnamefont {van
  Wezel}},\ and\ \bibinfo {author} {\bibfnamefont {C.}~\bibnamefont
  {Coulais}},\ }\bibfield  {title} {\bibinfo {title} {Observation of
  non-hermitian topology and its bulk-edge correspondence in an active
  mechanical metamaterial},\ }\href {https://doi.org/10.1073/pnas.2010580117}
  {\bibfield  {journal} {\bibinfo  {journal} {Proc. Natl. Acad. Sci. U.S.A.}\
  }\textbf {\bibinfo {volume} {117}},\ \bibinfo {pages} {29561} (\bibinfo
  {year} {2020})}\BibitemShut {NoStop}%
\bibitem [{\citenamefont {Xiao}\ \emph {et~al.}(2020)\citenamefont {Xiao},
  \citenamefont {Deng}, \citenamefont {Wang}, \citenamefont {Zhu},
  \citenamefont {Wang}, \citenamefont {Yi},\ and\ \citenamefont
  {Xue}}]{Nat.Phys.16.761}%
  \BibitemOpen
  \bibfield  {author} {\bibinfo {author} {\bibfnamefont {L.}~\bibnamefont
  {Xiao}}, \bibinfo {author} {\bibfnamefont {T.}~\bibnamefont {Deng}}, \bibinfo
  {author} {\bibfnamefont {K.}~\bibnamefont {Wang}}, \bibinfo {author}
  {\bibfnamefont {G.}~\bibnamefont {Zhu}}, \bibinfo {author} {\bibfnamefont
  {Z.}~\bibnamefont {Wang}}, \bibinfo {author} {\bibfnamefont {W.}~\bibnamefont
  {Yi}},\ and\ \bibinfo {author} {\bibfnamefont {P.}~\bibnamefont {Xue}},\
  }\bibfield  {title} {\bibinfo {title} {Non-hermitian bulk–boundary
  correspondence in quantum dynamics},\ }\href
  {https://doi.org/10.1038/s41567-020-0836-6} {\bibfield  {journal} {\bibinfo
  {journal} {Nat. Phys.}\ }\textbf {\bibinfo {volume} {16}},\ \bibinfo {pages}
  {761} (\bibinfo {year} {2020})}\BibitemShut {NoStop}%
\bibitem [{\citenamefont {Weidemann}\ \emph {et~al.}(2020)\citenamefont
  {Weidemann}, \citenamefont {Kremer}, \citenamefont {Helbig}, \citenamefont
  {Hofmann}, \citenamefont {Stegmaier}, \citenamefont {Greiter}, \citenamefont
  {Thomas},\ and\ \citenamefont {Szameit}}]{lightfunnel}%
  \BibitemOpen
  \bibfield  {author} {\bibinfo {author} {\bibfnamefont {S.}~\bibnamefont
  {Weidemann}}, \bibinfo {author} {\bibfnamefont {M.}~\bibnamefont {Kremer}},
  \bibinfo {author} {\bibfnamefont {M.}~\bibnamefont {Helbig}}, \bibinfo
  {author} {\bibfnamefont {T.}~\bibnamefont {Hofmann}}, \bibinfo {author}
  {\bibfnamefont {A.}~\bibnamefont {Stegmaier}}, \bibinfo {author}
  {\bibfnamefont {M.}~\bibnamefont {Greiter}}, \bibinfo {author} {\bibfnamefont
  {R.}~\bibnamefont {Thomas}},\ and\ \bibinfo {author} {\bibfnamefont
  {A.}~\bibnamefont {Szameit}},\ }\bibfield  {title} {\bibinfo {title}
  {Topological funneling of light},\ }\href@noop {} {\bibfield  {journal}
  {\bibinfo  {journal} {Science}\ }\textbf {\bibinfo {volume} {368}},\ \bibinfo
  {pages} {311} (\bibinfo {year} {2020})}\BibitemShut {NoStop}%
\bibitem [{\citenamefont {Liang}\ \emph {et~al.}(2022)\citenamefont {Liang},
  \citenamefont {Xie}, \citenamefont {Dong}, \citenamefont {Li}, \citenamefont
  {Li}, \citenamefont {Gadway}, \citenamefont {Yi},\ and\ \citenamefont
  {Yan}}]{PhysRevLett.129.070401}%
  \BibitemOpen
  \bibfield  {author} {\bibinfo {author} {\bibfnamefont {Q.}~\bibnamefont
  {Liang}}, \bibinfo {author} {\bibfnamefont {D.}~\bibnamefont {Xie}}, \bibinfo
  {author} {\bibfnamefont {Z.}~\bibnamefont {Dong}}, \bibinfo {author}
  {\bibfnamefont {H.}~\bibnamefont {Li}}, \bibinfo {author} {\bibfnamefont
  {H.}~\bibnamefont {Li}}, \bibinfo {author} {\bibfnamefont {B.}~\bibnamefont
  {Gadway}}, \bibinfo {author} {\bibfnamefont {W.}~\bibnamefont {Yi}},\ and\
  \bibinfo {author} {\bibfnamefont {B.}~\bibnamefont {Yan}},\ }\bibfield
  {title} {\bibinfo {title} {Dynamic signatures of non-hermitian skin effect
  and topology in ultracold atoms},\ }\href
  {https://doi.org/10.1103/PhysRevLett.129.070401} {\bibfield  {journal}
  {\bibinfo  {journal} {Phys. Rev. Lett.}\ }\textbf {\bibinfo {volume} {129}},\
  \bibinfo {pages} {070401} (\bibinfo {year} {2022})}\BibitemShut {NoStop}%
\bibitem [{\citenamefont {Xiao}\ \emph {et~al.}(2021)\citenamefont {Xiao},
  \citenamefont {Deng}, \citenamefont {Wang}, \citenamefont {Wang},
  \citenamefont {Yi},\ and\ \citenamefont {Xue}}]{nonblochep}%
  \BibitemOpen
  \bibfield  {author} {\bibinfo {author} {\bibfnamefont {L.}~\bibnamefont
  {Xiao}}, \bibinfo {author} {\bibfnamefont {T.}~\bibnamefont {Deng}}, \bibinfo
  {author} {\bibfnamefont {K.}~\bibnamefont {Wang}}, \bibinfo {author}
  {\bibfnamefont {Z.}~\bibnamefont {Wang}}, \bibinfo {author} {\bibfnamefont
  {W.}~\bibnamefont {Yi}},\ and\ \bibinfo {author} {\bibfnamefont
  {P.}~\bibnamefont {Xue}},\ }\bibfield  {title} {\bibinfo {title} {Observation
  of non-bloch parity-time symmetry and exceptional points},\ }\href
  {https://doi.org/10.1103/PhysRevLett.126.230402} {\bibfield  {journal}
  {\bibinfo  {journal} {Phys. Rev. Lett.}\ }\textbf {\bibinfo {volume} {126}},\
  \bibinfo {pages} {230402} (\bibinfo {year} {2021})}\BibitemShut {NoStop}%
\bibitem [{\citenamefont {Wang}\ \emph {et~al.}(2021)\citenamefont {Wang},
  \citenamefont {Li}, \citenamefont {Xiao}, \citenamefont {Han}, \citenamefont
  {Yi},\ and\ \citenamefont {Xue}}]{nonblochquench}%
  \BibitemOpen
  \bibfield  {author} {\bibinfo {author} {\bibfnamefont {K.}~\bibnamefont
  {Wang}}, \bibinfo {author} {\bibfnamefont {T.}~\bibnamefont {Li}}, \bibinfo
  {author} {\bibfnamefont {L.}~\bibnamefont {Xiao}}, \bibinfo {author}
  {\bibfnamefont {Y.}~\bibnamefont {Han}}, \bibinfo {author} {\bibfnamefont
  {W.}~\bibnamefont {Yi}},\ and\ \bibinfo {author} {\bibfnamefont
  {P.}~\bibnamefont {Xue}},\ }\bibfield  {title} {\bibinfo {title} {Detecting
  non-bloch topological invariants in quantum dynamics},\ }\href
  {https://doi.org/10.1103/PhysRevLett.127.270602} {\bibfield  {journal}
  {\bibinfo  {journal} {Phys. Rev. Lett.}\ }\textbf {\bibinfo {volume} {127}},\
  \bibinfo {pages} {270602} (\bibinfo {year} {2021})}\BibitemShut {NoStop}%
\bibitem [{\citenamefont {Gou}\ \emph {et~al.}(2020)\citenamefont {Gou},
  \citenamefont {Chen}, \citenamefont {Xie}, \citenamefont {Xiao},
  \citenamefont {Deng}, \citenamefont {Gadway}, \citenamefont {Yi},\ and\
  \citenamefont {Yan}}]{PhysRevLett.124.070402}%
  \BibitemOpen
  \bibfield  {author} {\bibinfo {author} {\bibfnamefont {W.}~\bibnamefont
  {Gou}}, \bibinfo {author} {\bibfnamefont {T.}~\bibnamefont {Chen}}, \bibinfo
  {author} {\bibfnamefont {D.}~\bibnamefont {Xie}}, \bibinfo {author}
  {\bibfnamefont {T.}~\bibnamefont {Xiao}}, \bibinfo {author} {\bibfnamefont
  {T.-S.}\ \bibnamefont {Deng}}, \bibinfo {author} {\bibfnamefont
  {B.}~\bibnamefont {Gadway}}, \bibinfo {author} {\bibfnamefont
  {W.}~\bibnamefont {Yi}},\ and\ \bibinfo {author} {\bibfnamefont
  {B.}~\bibnamefont {Yan}},\ }\bibfield  {title} {\bibinfo {title} {Tunable
  nonreciprocal quantum transport through a dissipative aharonov-bohm ring in
  ultracold atoms},\ }\href {https://doi.org/10.1103/PhysRevLett.124.070402}
  {\bibfield  {journal} {\bibinfo  {journal} {Phys. Rev. Lett.}\ }\textbf
  {\bibinfo {volume} {124}},\ \bibinfo {pages} {070402} (\bibinfo {year}
  {2020})}\BibitemShut {NoStop}%
\bibitem [{\citenamefont {Blatt}\ and\ \citenamefont
  {Roos}(2012)}]{doi.org/10.1038/nphys2252}%
  \BibitemOpen
  \bibfield  {author} {\bibinfo {author} {\bibfnamefont {R.}~\bibnamefont
  {Blatt}}\ and\ \bibinfo {author} {\bibfnamefont {C.~F.}\ \bibnamefont
  {Roos}},\ }\bibfield  {title} {\bibinfo {title} {Quantum simulations with
  trapped ions},\ }\href {https://doi.org/10.1038/nphys2252} {\bibfield
  {journal} {\bibinfo  {journal} {Nat. Phys.}\ }\textbf {\bibinfo {volume}
  {8}},\ \bibinfo {pages} {277} (\bibinfo {year} {2012})}\BibitemShut {NoStop}%
\bibitem [{\citenamefont {Barreiro}\ \emph {et~al.}(2011)\citenamefont
  {Barreiro}, \citenamefont {Müller}, \citenamefont {Schindler}, \citenamefont
  {Nigg}, \citenamefont {Monz}, \citenamefont {Chwalla}, \citenamefont
  {Hennrich}, \citenamefont {Roos}, \citenamefont {Zoller},\ and\ \citenamefont
  {Blatt}}]{doi.org/10.1038/nature09801}%
  \BibitemOpen
  \bibfield  {author} {\bibinfo {author} {\bibfnamefont {J.~T.}\ \bibnamefont
  {Barreiro}}, \bibinfo {author} {\bibfnamefont {M.}~\bibnamefont {Müller}},
  \bibinfo {author} {\bibfnamefont {P.}~\bibnamefont {Schindler}}, \bibinfo
  {author} {\bibfnamefont {D.}~\bibnamefont {Nigg}}, \bibinfo {author}
  {\bibfnamefont {T.}~\bibnamefont {Monz}}, \bibinfo {author} {\bibfnamefont
  {M.}~\bibnamefont {Chwalla}}, \bibinfo {author} {\bibfnamefont
  {M.}~\bibnamefont {Hennrich}}, \bibinfo {author} {\bibfnamefont {C.~F.}\
  \bibnamefont {Roos}}, \bibinfo {author} {\bibfnamefont {P.}~\bibnamefont
  {Zoller}},\ and\ \bibinfo {author} {\bibfnamefont {R.}~\bibnamefont
  {Blatt}},\ }\bibfield  {title} {\bibinfo {title} {An open-system quantum
  simulator with trapped ions},\ }\href {https://doi.org/10.1038/nature09801}
  {\bibfield  {journal} {\bibinfo  {journal} {Nature}\ }\textbf {\bibinfo
  {volume} {470}},\ \bibinfo {pages} {486} (\bibinfo {year}
  {2011})}\BibitemShut {NoStop}%
\bibitem [{\citenamefont {Wannier}(1960)}]{PhysRev.117.432}%
  \BibitemOpen
  \bibfield  {author} {\bibinfo {author} {\bibfnamefont {G.~H.}\ \bibnamefont
  {Wannier}},\ }\bibfield  {title} {\bibinfo {title} {Wave functions and
  effective hamiltonian for bloch electrons in an electric field},\ }\href
  {https://doi.org/10.1103/PhysRev.117.432} {\bibfield  {journal} {\bibinfo
  {journal} {Phys. Rev.}\ }\textbf {\bibinfo {volume} {117}},\ \bibinfo {pages}
  {432} (\bibinfo {year} {1960})}\BibitemShut {NoStop}%
\bibitem [{\citenamefont {Cai}\ \emph {et~al.}(2021)\citenamefont {Cai},
  \citenamefont {Liu}, \citenamefont {Zhao}, \citenamefont {Wu}, \citenamefont
  {Mei}, \citenamefont {Jiang}, \citenamefont {He}, \citenamefont {Zhang},
  \citenamefont {Zhou},\ and\ \citenamefont
  {Duan}}]{Nature.Communications.12.1126.(2021)}%
  \BibitemOpen
  \bibfield  {author} {\bibinfo {author} {\bibfnamefont {M.~L.}\ \bibnamefont
  {Cai}}, \bibinfo {author} {\bibfnamefont {Z.~D.}\ \bibnamefont {Liu}},
  \bibinfo {author} {\bibfnamefont {W.~D.}\ \bibnamefont {Zhao}}, \bibinfo
  {author} {\bibfnamefont {Y.~K.}\ \bibnamefont {Wu}}, \bibinfo {author}
  {\bibfnamefont {Q.~X.}\ \bibnamefont {Mei}}, \bibinfo {author} {\bibfnamefont
  {Y.}~\bibnamefont {Jiang}}, \bibinfo {author} {\bibfnamefont
  {L.}~\bibnamefont {He}}, \bibinfo {author} {\bibfnamefont {X.}~\bibnamefont
  {Zhang}}, \bibinfo {author} {\bibfnamefont {Z.~C.}\ \bibnamefont {Zhou}},\
  and\ \bibinfo {author} {\bibfnamefont {L.~M.}\ \bibnamefont {Duan}},\
  }\bibfield  {title} {\bibinfo {title} {Observation of a quantum phase
  transition in the quantum rabi model with a single trapped ion},\ }\href
  {https://doi.org/10.1038/s41467-021-21425-8} {\bibfield  {journal} {\bibinfo
  {journal} {Nat. Commun.}\ }\textbf {\bibinfo {volume} {12}},\ \bibinfo
  {pages} {1126} (\bibinfo {year} {2021})}\BibitemShut {NoStop}%
\bibitem [{\citenamefont {McCormick}\ \emph {et~al.}(2019)\citenamefont
  {McCormick}, \citenamefont {Keller}, \citenamefont {Burd}, \citenamefont
  {Wineland}, \citenamefont {Wilson},\ and\ \citenamefont
  {Leibfried}}]{Nature.572.86(2019)}%
  \BibitemOpen
  \bibfield  {author} {\bibinfo {author} {\bibfnamefont {K.~C.}\ \bibnamefont
  {McCormick}}, \bibinfo {author} {\bibfnamefont {J.}~\bibnamefont {Keller}},
  \bibinfo {author} {\bibfnamefont {S.~C.}\ \bibnamefont {Burd}}, \bibinfo
  {author} {\bibfnamefont {D.~J.}\ \bibnamefont {Wineland}}, \bibinfo {author}
  {\bibfnamefont {A.~C.}\ \bibnamefont {Wilson}},\ and\ \bibinfo {author}
  {\bibfnamefont {D.}~\bibnamefont {Leibfried}},\ }\bibfield  {title} {\bibinfo
  {title} {Quantum-enhanced sensing of a single-ion mechanical oscillator},\
  }\href {https://doi.org/10.1038/s41586-019-1421-y} {\bibfield  {journal}
  {\bibinfo  {journal} {Nature}\ }\textbf {\bibinfo {volume} {572}},\ \bibinfo
  {pages} {86} (\bibinfo {year} {2019})}\BibitemShut {NoStop}%
\bibitem [{\citenamefont {Bloch}(1929)}]{BO}%
  \BibitemOpen
  \bibfield  {author} {\bibinfo {author} {\bibfnamefont {F.}~\bibnamefont
  {Bloch}},\ }\bibfield  {title} {\bibinfo {title} {Über die quantenmechanik
  der elektronen in kristallgittern},\ }\href
  {https://doi.org/10.1007/BF01339455} {\bibfield  {journal} {\bibinfo
  {journal} {Z. Phys.}\ }\textbf {\bibinfo {volume} {52}},\ \bibinfo {pages}
  {555} (\bibinfo {year} {1929})}\BibitemShut {NoStop}%
\bibitem [{\citenamefont {Li}\ \emph {et~al.}(2022)\citenamefont {Li},
  \citenamefont {Zhang},\ and\ \citenamefont {Yi}}]{tianyuprb}%
  \BibitemOpen
  \bibfield  {author} {\bibinfo {author} {\bibfnamefont {T.}~\bibnamefont
  {Li}}, \bibinfo {author} {\bibfnamefont {Y.-S.}\ \bibnamefont {Zhang}},\ and\
  \bibinfo {author} {\bibfnamefont {W.}~\bibnamefont {Yi}},\ }\bibfield
  {title} {\bibinfo {title} {Engineering dissipative quasicrystals},\
  }\href@noop {} {\bibfield  {journal} {\bibinfo  {journal} {Phys. Rev. B}\
  }\textbf {\bibinfo {volume} {105}},\ \bibinfo {pages} {125111} (\bibinfo
  {year} {2022})}\BibitemShut {NoStop}%
\bibitem [{\citenamefont {Zhang}\ \emph {et~al.}(2022)\citenamefont {Zhang},
  \citenamefont {Yang},\ and\ \citenamefont
  {Fang}}]{doi.org/10.1038/s41467-022-30161-6}%
  \BibitemOpen
  \bibfield  {author} {\bibinfo {author} {\bibfnamefont {K.}~\bibnamefont
  {Zhang}}, \bibinfo {author} {\bibfnamefont {Z.}~\bibnamefont {Yang}},\ and\
  \bibinfo {author} {\bibfnamefont {C.}~\bibnamefont {Fang}},\ }\bibfield
  {title} {\bibinfo {title} {Universal non-hermitian skin effect in two and
  higher dimensions},\ }\href {https://doi.org/10.1038/s41467-022-30161-6}
  {\bibfield  {journal} {\bibinfo  {journal} {Nat. Commun.}\ }\textbf {\bibinfo
  {volume} {13}},\ \bibinfo {pages} {2496} (\bibinfo {year}
  {2022})}\BibitemShut {NoStop}%
\bibitem [{\citenamefont {Monroe}\ \emph {et~al.}(2021)\citenamefont {Monroe},
  \citenamefont {Campbell}, \citenamefont {Duan}, \citenamefont {Gong},
  \citenamefont {Gorshkov}, \citenamefont {Hess}, \citenamefont {Islam},
  \citenamefont {Kim}, \citenamefont {Linke}, \citenamefont {Pagano},
  \citenamefont {Richerme}, \citenamefont {Senko},\ and\ \citenamefont
  {Yao}}]{RevModPhys.93.025001}%
  \BibitemOpen
  \bibfield  {author} {\bibinfo {author} {\bibfnamefont {C.}~\bibnamefont
  {Monroe}}, \bibinfo {author} {\bibfnamefont {W.~C.}\ \bibnamefont
  {Campbell}}, \bibinfo {author} {\bibfnamefont {L.-M.}\ \bibnamefont {Duan}},
  \bibinfo {author} {\bibfnamefont {Z.-X.}\ \bibnamefont {Gong}}, \bibinfo
  {author} {\bibfnamefont {A.~V.}\ \bibnamefont {Gorshkov}}, \bibinfo {author}
  {\bibfnamefont {P.~W.}\ \bibnamefont {Hess}}, \bibinfo {author}
  {\bibfnamefont {R.}~\bibnamefont {Islam}}, \bibinfo {author} {\bibfnamefont
  {K.}~\bibnamefont {Kim}}, \bibinfo {author} {\bibfnamefont {N.~M.}\
  \bibnamefont {Linke}}, \bibinfo {author} {\bibfnamefont {G.}~\bibnamefont
  {Pagano}}, \bibinfo {author} {\bibfnamefont {P.}~\bibnamefont {Richerme}},
  \bibinfo {author} {\bibfnamefont {C.}~\bibnamefont {Senko}},\ and\ \bibinfo
  {author} {\bibfnamefont {N.~Y.}\ \bibnamefont {Yao}},\ }\bibfield  {title}
  {\bibinfo {title} {Programmable quantum simulations of spin systems with
  trapped ions},\ }\href {https://doi.org/10.1103/RevModPhys.93.025001}
  {\bibfield  {journal} {\bibinfo  {journal} {Rev. Mod. Phys.}\ }\textbf
  {\bibinfo {volume} {93}},\ \bibinfo {pages} {025001} (\bibinfo {year}
  {2021})}\BibitemShut {NoStop}%
\end{thebibliography}

\providecommand{\noopsort}[1]{}\providecommand{\singleletter}[1]{#1}%

\end{document}